\documentclass[12pt,preprint]{aastex}
\usepackage{graphicx}
\usepackage{emulateapj5}
\usepackage{apjfonts}
\usepackage{epsfig}
\usepackage{natbib}

\newcommand{\brg}{Br\ensuremath{\gamma}}

\def\gtrsim{\mathrel{\hbox{\rlap{\hbox{\lower4pt\hbox{$\sim$}}}\hbox{\raise2pt\hbox{$>$}}}}}

\newcommand{\halpha}{H\ensuremath{\alpha}}

\newcommand{\hmol}{H\ensuremath{_2}}

\newcommand{\hst}{\emph{HST}}

\newcommand{\kms}{km~s\ensuremath{^{-1}}}

\newcommand{\lledd}{\ensuremath{L_{\mathrm{bol}}/L{\mathrm{_{Edd}}}}}

\newcommand{\msun}{\ensuremath{M_{\odot}}}

\newcommand{\oiii}{[\ion{O}{3}]}

\newcommand{\sivi}{[Si {\small VI}]}

\newcommand{\sigmastar}{\ensuremath{\sigma_{\ast}}}

\def\lax{{$\mathrel{\hbox{\rlap{\hbox{\lower4pt\hbox{$\sim$}}}\hbox{$<$}}}$}}
\def\gax{{$\mathrel{\hbox{\rlap{\hbox{\lower4pt\hbox{$\sim$}}}\hbox{$>$}}}$}}

\slugcomment{Accepted version; May 4, 2014;
to be published in {\it The Astrophysical Journal}.}
\shorttitle{{\it Molecular Gas in Megamaser Galaxies}}
\shortauthors{GREENE, ET AL.}

\begin{document}

\title{Circumnuclear Molecular Gas in Megamaser Disk Galaxies 
NGC 4388 and NGC 1194}

\author{Jenny E. Greene\altaffilmark{1}, 
Anil Seth\altaffilmark{2}, Mariya Lyubenova\altaffilmark{3}, 
Jonelle Walsh\altaffilmark{4}, Glenn van de Ven\altaffilmark{3},
Ronald L{\"a}sker\altaffilmark{3}}

\altaffiltext{1}{Department of Astrophysics, Princeton University, Princeton, NJ 08540}
\altaffiltext{2}{University of Utah, Salt Lake City, UT 84112}
\altaffiltext{3}{Max Planck Institute for Astronomy, K\"onigstuhl 17, 69117 
Heidelberg, Germany}
\altaffiltext{4}{Department of Astronomy, The University of Texas at Austin, 
2515 Speedway, Stop C1400, Austin, TX 78712-1205, USA}

\begin{abstract}
  We explore the warm molecular and ionized gas in the centers of two
  megamaser disk galaxies using $K-$band spectroscopy. Our ultimate
  goal is to determine how gas is funneled onto the accretion disk, here
  traced by megamaser spots on sub-pc scales.  We present NIR IFU data
  with a resolution of $\sim 50$ pc for two galaxies: NGC 4388 with
  VLT/SINFONI and NGC 1194 with Keck/OSIRIS+AO.  The high spatial
  resolution and rich spectral diagnostics allow us to study both the
  stellar and gas kinematics as well as gas excitation on scales only
  an order of magnitude larger than the maser disk.  We find a drop in
  the stellar velocity dispersion in the inner $\sim 100$ pc of NGC
  4388, a common signature of a dynamically cold central component
  seen in many active nuclei. We also see evidence for non-circular
  gas motions in the molecular hydrogen on similar scales, with the
  gas kinematics on 100-pc scales aligned with the megamaser disk. In
  contrast, the high ionization lines and \brg\ trace outflow along
  the 100 pc-scale jet. In NGC 1194, the continuum from the accreting
  black hole is very strong, making it difficult to measure robust
  two-dimensional kinematics, but the spatial distribution and line
  ratios of the molecular hydrogen and Br$\gamma$ have consistent
  properties between the two galaxies.
\end{abstract}

\section{Introduction}
\label{sec:Introduction}

Active galactic nuclei pose a fundamental problem -- how does rotationally dominated 
gas on kpc scales lose orders of magnitude in angular momentum
to fall onto an accretion disk on AU scales
\citep[e.g.,][]{balickheckman1982}.  We do not know the mechanism that
dissipates angular momentum and allows gas to accrete.  There are no
shortage of ideas, including major or minor mergers
\citep[e.g.,][]{dahari1985,mihoshernquist1994,kuoetal2008,ellisonetal2011},
bars and bars within bars
\citep[e.g.,][]{shlosmanetal1990,maciejewskietal2002,
  huntetal2008,kimetal2012}, and nuclear spirals
\citep[][]{englmaiershlosman2000,maciejewski2004,
  martinietal2003,annthakur2005} as perhaps indicated by dust lanes
\citep{simoeslopesetal2007,martinietal2013}.  Even looking directly at
molecular gas kinematics on pc scales in active galaxies, it is very
hard to find clear evidence for the true driver of nuclear activity
\citep[][]{haanetal2009,garciaburilloetal2009,garciaburillocombes2012,
  combesetal2014}.

Thanks to the advent of near-infrared integral-field spectrographs
(NIR IFU hereafter), there has been significant progress in
understanding gas flows in the centers of active galactic nuclei (AGN)
in recent years.  The NIR observations allow us to penetrate 
gas and dust in the active nuclei, and, in conjunction with 
adaptive optics, to probe very near to the nucleus. 
In some nearby cases, inflows are directly
observed along circumnuclear spirals in ionized gas on $\sim 100$ pc
scales \citep[e.g.,][]{storchibergmannetal2007,daviesetal2009,vandevenfathi2010,
  riffeletal2013}.  Furthermore, there are intriguing
hints of kinematic differences in the nuclei of local Seyfert
galaxies.  For example, \citet{dumasetal2007} suggest that the ionized
gas in the disks of Seyfert galaxies (i.e., outside of the narrow-line
region or NLR) is more kinematically disturbed than the gas in a
control inactive sample.  \citet{hicksetal2013} also report that
Seyfert galaxies have more concentrated stellar luminosity profiles, 
lower stellar velocity dispersions, and elevated \hmol\ 1-0 S(1) 
emission within $\sim 100-200$ pc relative to an inactive subsample.

In this work, we specifically focus on megamaser disk galaxies.  In
these special systems, luminous water megamasers at 22 GHz trace a
geometrically thin molecular disk on sub-pc scales in orbit around the
central BH
\citep[e.g.,][]{miyoshietal1995,herrnsteinetal2005,lo2005,kuoetal2011}.
The megamaser spots show near perfect Keplerian rotation in many
\citep[e.g.,][]{kuoetal2011}, but not all cases
\citep[e.g.,][]{lodatobertin2003,kondratkoetal2005}.  Thanks to the
Keplerian rotation in the maser disk, we know the BH mass to a few
percent, limited in most cases by the distance to the galaxy
\citep[e.g.,][]{kuoetal2011}.  Direct geometric distances have also
been derived by measuring the accelerations of the systemic masers for
the few best cases
\citep{herrnsteinetal2005,reidetal2009,reidetal2013, kuoetal2013}.

In addition to distance measurements and BH masses, the megamaser disk
reveals the spin axis of the accretion disk on sub-pc scales.  To get
the long path lengths required for masing, the disk must be virtually
edge-on; we thus typically know the orientation of the disk on sub-pc
scales to within a couple of degrees \citep{kuoetal2011}.  We have
already found that the spin axis of the megamaser disk aligns with the
jet on $\sim 100$ pc scales, while the spin axis of the disk does not
appear to align with flattened disk-like structures on $\sim 500$ pc
scales identified in \hst\ imaging \citep{greeneetal2013}.  In this
paper, we explore the use of NIR IFU data to define the gas flows on
$\lesssim 500$~pc scales, for comparison with the megamaser disk.  We
currently have NIR IFU observations of the centers of two galaxies in
the Kuo et al.\ sample: NGC 4388 and NGC 1194.  These two are the
  only two megamaser disk galaxies with published BH masses where we
  can come close to resolving the gravitational sphere of influence,
  in the former case because the galaxy is nearby (in Virgo) and in
  the latter because the BH is the most massive known with a maser
  disk \citep{kuoetal2011}.

\section{Observations and Data Reduction}
\label{sec:Observations}

\hbox{
\hskip -4mm
\includegraphics[width=0.45\textwidth]{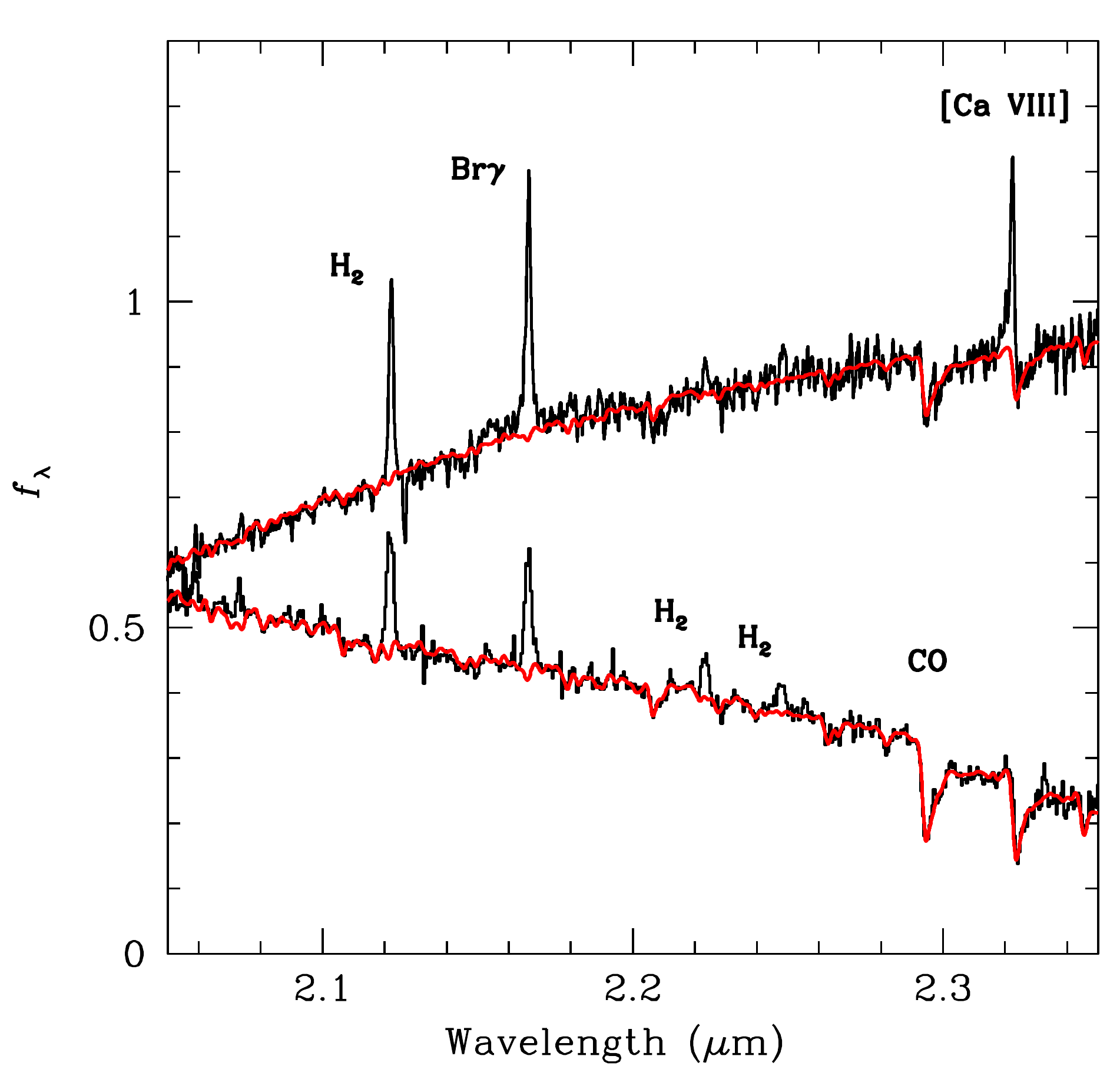}
}
\figcaption{Example fits to the stellar continuum of NGC 4388 using pPXF and
  stellar templates observed with SINFONI.  The top spectrum is the
  sum of pixels within the AGN-dominated central point source, showing
  the strong nonthermal AGN continuum. The bottom spectrum represents
  the sum of all high S/N pixels from the galaxy disk excluding the central 
  point source, which we fitted to derive a best-fit average template, as
  described in \S \ref{sec:stellarkin4388}.  The relatively blue
  spectrum here is dominated by stellar light.  Due to tellurice residuals, 
  we truncate the spectra at the blue end for display purposes.
\label{fig:specn4388}
}
\vskip 5mm

\subsection{NGC 4388 with SINFONI}

We observed NGC 4388 with SINFONI under program 
386.B-0244A\footnote{We are grateful to the
  ESO astronomers who obtained the data presented in this paper in
  service mode operations at La Silla Paranal Observatory.} on the
Very Large Telescope \citep{eis03,bonnet04} on April 12 \& May 5 2011
in natural seeing conditions and in service mode. We have been awarded
time to observe this galaxy at two spatial resolutions, the higher one
involving assistance of the Adaptive Optics system (AO), however to
date these observations are only partially carried out. Thus, in this
paper we present the lower spatial resolution data set that covers a
wider field of view.

Our observations cover the central 8\arcsec $\times$ 8\arcsec, with a
spatial sampling of 0\farcs125 $\times$ 0\farcs250. We used the
$K$-band grating (1.95 -- 2.45 $\mu$m) that gives a spectral
resolution R$\sim$3500 (6.2 \AA\ FWHM, as measured on sky lines). For
the observations we used standard near-IR nodding techniques. 
Observing blocks consisted of a sequence of object and sky frames
(OSOOSOOSOOSOO). Each individual integration was 300 sec and the sky
fields were offset by 240\arcsec\ to the North and East. Science frames
were dithered by 0\farcs3 in order to reject bad pixels. There were
two observing blocks,  with a total on-source integration time of 100 min,
and 40 min in total on sky frames. Additionally,
after each observing block and at a similar airmass, we observed a B
dwarf star to act as a telluric star.

We used the ESO SINFONI pipeline v2.3.3 to perform the basic data
reduction on each observing block, consisting of 10 object and 4 sky
exposures. In brief, the pipeline extracts the raw data, applies
distortion, bad pixel and flat-field corrections and wavelength
calibration, and stores the combined sky-subtracted spectra from one
observing block in a 3-dimensional data cube. The final pixel size is 
$0\farcs125$.

We reduced the telluric stars in the same way as the science
frames. For each telluric star we extracted a one-dimensional
spectrum, removed the hydrogen Brackett\,$\gamma$ absorption line at
$2.166\,\mu$m after fitting it with a Lorentzian profile, and divided
the star spectrum by a black body spectrum with the same temperature
as the star. The last step in preparing the telluric spectrum was to
apply small shifts in the spectral direction ($<$0.05 pixels or
0.123 \AA) and scalings to minimize the residuals of the telluric
features. To do this, we extracted a central one-dimensional spectrum
from each science data cube and cross-correlated and fitted it with
the corresponding telluric spectrum. We derived the
wavelength-dependent correction required to match the continuum shape
of the telluric star (without an absolute zero-point).  Thus, when we
divided each individual spaxel in the six galaxy data cubes by the
corresponding best-fitting telluric spectrum, we also obtained a
relative flux calibration.

Based on the seeing monitor, during our observations the seeing varied
from $1-1.5$\arcsec\ for the first observation, and between
$0.5-0.7$\arcsec\ for the second observation. Note that the seeing
monitor takes measurements at 5000~\AA, thus our seeing is typically
better than this. From the standard stars (taken at the end) we
measure a FWHM of 0\farcs8 for the first observation and 0\farcs5 for
the second.  We aligned the two OBs with integer offsets and summed
them; all analysis was done on this final summed data cube.

\subsection{NGC 1194 with OSIRIS}

We observed NGC 1194 using the IFU OH-Suppressing Infrared Imaging
Spectrograph \citep[OSIRIS][]{larkinetal2006} assisted by the
laser-guide star adaptive optics system
\citep{wizinowichetal2006,vandametal2006} on the 10m Keck II
telescope.  The data were taken over the latter two-thirds of a single
night on Oct.\ 19 2011.  We observed in the $K-$band with the Kbb
filter, for a spectral coverage of $1.965-2.381~\micron$ and a
spectral resolution of $R \approx 4000$. We used the $0\farcs05$
spatial scale, and oriented the long axis of the IFU along the major
axis of the galaxy at a PA of $145$\degr.  We dithered the galaxy
by $0\farcs2$ perpendicular to the long axis of the IFU, both to
facilitate bad pixel removal and to slightly increase the
field-of-view to $1\farcs2 \times 3 \farcs2$.

Unfortunately, the nucleus of NGC 1194 was too faint to use as a
tip-tilt star.  Thus, we used a nearby (54\arcsec) star
(ID=0888-0029937; $R=14.7$).  Natural seeing conditions varied over
the night from $0\farcs5$ to $1\farcs1$.  From our final science data
cube we find that the PSF core had a FWHM of $0\farcs16$ 
as measured from the galaxy core.  We also
observed a telluric standard (A0V) star approximately once per unit
change in airmass. Finally, we interleaved sky and object observations
with an Object-Sky-Object pattern, spending 600s per sky and object
exposure. We acquired a total of 170 min on source.

We followed closely the reductions outlined in \citet{walshetal2012},
using the OSIRIS data reduction pipeline (v2.3) made available by the
instrument
team\footnote{http://irlab.astro.ucla.edu/osiris/pipeline.html}.  We
repeat salient details here for completeness.  The pipeline handles
sky subtraction, cosmic-ray removal, glitch identification, spectral
extraction into a data cube, wavelength calibration, atmospheric
dispersion correction, and telluric correction using an A0V star.  We
experimented with a scaled sky subtraction method from
\citet{davies2007}, but did not find substantial improvements from the
direct method.  Finally, we determined the centroid of each exposure
by hand, applied these sub-pixel offsets, and averaged all cubes into
our final data cube.

\section{Spectral Analysis}

Here we describe our analysis procedure for the NGC 4388-SINFONI data
cube in detail, highlighting any differences in analysis with the
NGC 1194 data cube.

\subsection{Stellar Kinematics}

We use the direct-pixel--fitting code pPXF of
\citep{cappellariemsellem2004} to fit the stellar continuum.  To
achieve adequate signal-to-noise (S/N) we use the Voronoi binning
scheme of \citet{cappellaricopin2003}, which determines contiguous
bins with uniform S/N, in this case S/N$>50$ per pixel.  The SINFONI
pipeline does not return an error array, and so the noise is
determined empirically from line-free regions of the spectrum as the
variance in the spectrum once large outliers are clipped. In the case
of the OSIRIS data, we use the per-pixel error array generated by the
reduction pipeline. We then fit the coadded spectrum corresponding to
each bin with a weighted combination of stellar templates.  Each
template is shifted to the systemic velocity of the galaxy and
convolved with a Gaussian line-broadening function. A polynomial of
fourth order is added to account for nonthermal continuum from the
AGN, and potentially flux calibration errors as well.  Example fits
are shown in Figure \ref{fig:specn4388}.

The measurement uncertainties in both velocity and velocity dispersion
are determined with Monte Carlo simulations.  In the case of the NGC
1194 data, the data reduction software generates an error array.
However, for the NGC 4388 data, we use residuals from the best fit
(with the emission lines removed) to determine the average sigma.  We
create 100 perturbed input spectra assuming Gaussian errors in the
spectra. We then refit these 100 artificial spectra.  The uncertainty
in each parameter is calculated as the values encompassing 68\% of the
trials.

We experimented with a number of fitting regions, including a fit to
the bandhead only ($2.2-2.4~\micron$, short), a fit to the full region
($1.95-2.4~\micron$, full), and an intermediate region ($2.04-2.4
~\micron$, best).  We found that the first fit delivered reliable
velocity dispersions, but unstable radial velocities because of the
narrow wavelength range, and because the bandhead at $2.32~\micron$ is
often filled in by [\ion{Ca}{8}]$~\lambda 2.322 ~\micron$ emission.
Fits to the full spectral region suffered from telluric residuals at
$\sim 2 ~\micron$.  The just-right spectral region was therefore the
third, intermediate case. The median errors on the velocities are $7,
12, 9.5$~\kms\ for the best, short, and full regions respectively,
while the median errors on the dispersion measurements are $8.5, 8.5,
12$~\kms.  In all cases, we masked high-EW emission lines, including
the \hmol\ lines at $2.034,\, 2.042,\, 2.066,\, 2.073,\, 2.122,\,
2.154,\, 2.201,\, 2.211,\, 2.223$ $,\, 2.248,\, 2.254,\, 2.287,\,
2.345 ~\micron$ and also [\ion{Si}{6}]$~\lambda1.963 ~\micron$,
\ion{He}{1}$~\lambda 2.059 ~\micron$, Br$\gamma~\lambda 2.166
~\micron$, and [\ion{Ca}{8}].

To bracket uncertainties caused by template mismatch, we utilize two
stellar template libraries in fitting the SINFONI data.  One is
a library of stars observed with SINFONI using the same observational
set-up.  In this case, no correction for instrumental resolution is
required. The native SINFONI templates range from K4III to M5III (plus
a K4.5V star) \citep{lyubenovaetal2008}.  As
a check, we also use the stellar template library of
\citet{wallacehinkle1996}, observed at higher spectral resolution.  We
use this second template set only to test our sensitivity to template
mismatch.  The Wallace templates cover a wider range in spectral type,
including KM supergiants and KM giants. We discuss the 
stellar template fits to the NGC 4388 data in detail in S 4.1. 

Even with different templates, we find good agreement in the two sets
of stellar velocity dispersion measurements.  Taking $\sigma_{\rm S}$
as the answer based on the SINFONI templates and $\sigma_{\rm W}$ as
the same for the Wallace templates, we find $\langle (\sigma_{\rm
  S}-\sigma_{\rm W})/\sigma_{\rm S} \rangle = -0.05 \pm 0.2$ (where
the latter number is simply the standard deviation in this ratio).  We
find that the two measurements agree within 20\% (i.e., within the
observational uncertainties) and we find no significant systematic
offset.  In the case of the OSIRIS data, our primary fitting uses
  a single K5III star, which provides an acceptable fit to our
  moderate S/N data.

\subsection{Finding the Center}

\begin{figure*}
\hbox{
\hskip +20mm
\includegraphics[width=0.75\textwidth]{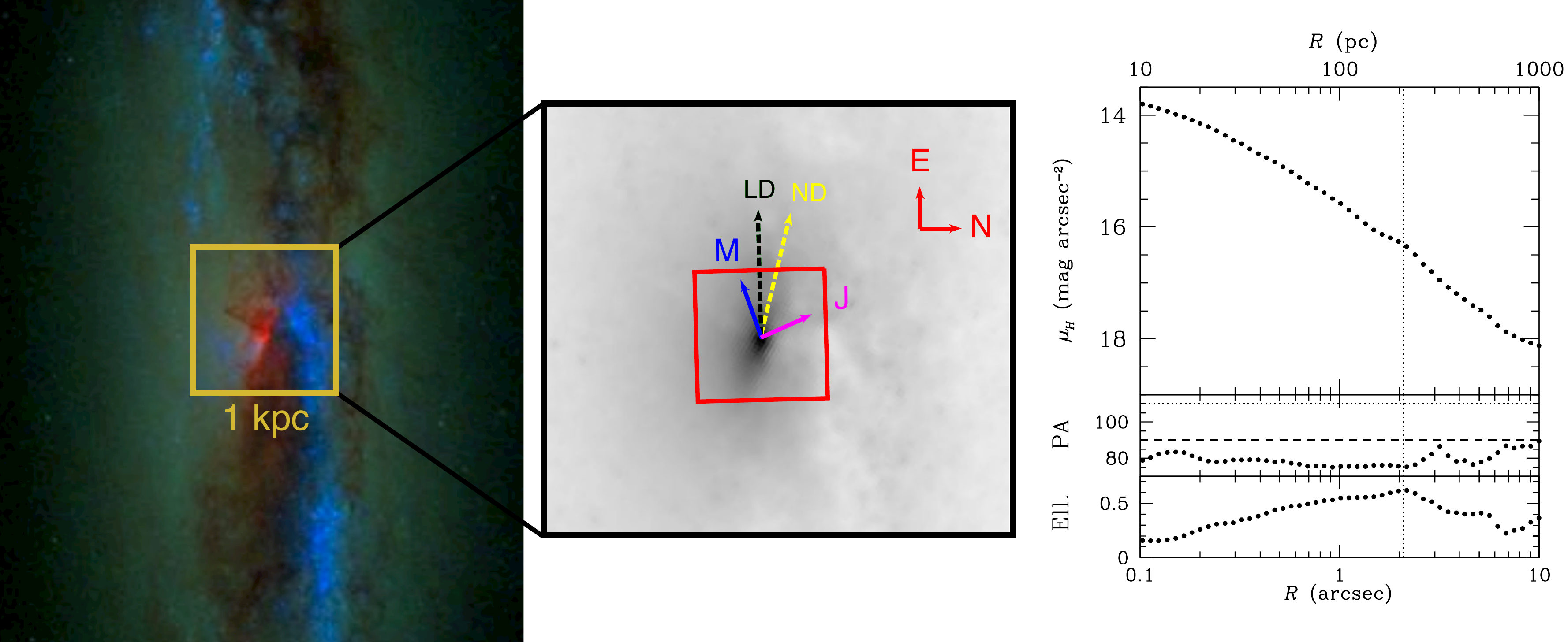}
}
\figcaption{{\it Left}: 
Three-band \hst/WFC3 image including filters F336W, F438W, and F814W
\citep{greeneetal2013}.  To match the data cube, E is up and N to the
right, identical to the middle image. The yellow box has a size of
11\arcsec\ ($\sim 1$ kpc), and matches the region shown in the
middle image.  {\it Middle}: A zoom-in on the nuclear disk as
revealed by our \hst/WFC3 F160W image.  The red box shows a $4\times
4$\arcsec\ region (360$\times$360 pc), corresponding to the SINFONI
IFU region that we display in all subsequent figures.  We
schematically indicate the 24\degr\ orientation of the 100 pc-scale
jet (J; magenta), the major-axis PA of the maser disk 
at 107\degr\ (M; blue), the large-scale
disk at 90\degr\ (LD; black dashed), and the nuclear disk at 75\degr\
(ND; yellow dashed). {\it Right}: Radial profile of the F160W image, 
including PA measured E of N, and ellipticity.  The
extent of the nuclear disk, with a PA of 75\degr, is indicated with the vertical
dotted lines. We also indicate with horizontal lines the PA of the
large-scale disk (90\degr, dashed) and the megamaser disk on sub-pc
scales (107\degr, dotted).
\label{fig:image}
}
\end{figure*}

We do not a priori know the location of the black hole, or the precise
photometric center of the galaxy.  We could use the continuum to
determine the photometric center, but then we would be sensitive to
obscuration, which is still significant in the NIR (see Figure
\ref{fig:image}). Instead, we use the equivalent width of the CO
bandhead to determine where the active nucleus peaks
\citep{daviesetal2004,daviesetal2007}.  We use the index definition
from \citet[][]{olivaetal1995}: the index band is
$2.2924-2.2977~\micron$, with a continuum band centered on 2.2900 and
a width of 0.0003~$\micron$, and we also followed their prescription
to correct for velocity dispersion although this correction is
negligible \citep[see also ][]{forsterschreiber2000}.  These authors
show that in the absence of an active nucleus, star-forming galaxies
have a very uniform CO EW. A declining CO EW towards the galaxy center
can be attributed to infill by the nonthermal continuum from the
accreting black hole.  We show an example in \S
\ref{sec:stellarkin4388}.

We create an AGN continuum map by taking the CO EW at the edge of the
cube (EW=9.5\AA\ in NGC 4388, consistent with stellar-population
models) and assuming that all dilution further inwards results from
the AGN, such that continuum (AGN) = [total continuum] $\times$ [1 -
CO EW/CO Outer]. The photometric center of the continuum map falls
$0\farcs0875$ South of the CO EW center.  The optical narrow-line
region cone is only visible to the South (\S \ref{sec:nlr}), so it is
sensible that the continuum centroid should be shifted towards the
low-reddening side of the disk.  The best-fit center also agrees
within less than a pixel with the peak \hmol\ and \brg\ emission,
although these latter are not as well defined.  We will use this
position as the center of the cube throughout. In the case of NGC
1194, the CO-derived and photo-centers agree within $< 0.005$\arcsec.

\subsection{Emission Line Fitting} 

While Voronoi binning is very useful to create uniform continuum S/N 
\citep[e.g.,][]{cidfernandesetal2013},
it is more difficult to apply to patchy emission-line maps, where 
by using large bins we may smear out
interesting features in the line emission maps.  We therefore create a second data
cube using a $3 \times 3$ pixel smoothing in the outer regions and a
$2 \times 2$ pixel binning in the inner $2 \times 2$\arcsec\
region. With this level of binning, we increase the S/N by a factor of 
two but still maintain multiple pixels across the PSF. 
We measure the emission lines from this uniformly gridded
cube.  In the case of NGC 4388, because we have such high S/N in the
continuum, we again perform continuum-subtraction using pPXF before
fitting the emission lines.  However, there are no strong stellar absorption
features beneath the emission lines of interest, and thus for
NGC 1194, where the S/N is poor in the continuum, we simply fit a 
local continuum value.

We fit each emission line independently. We model the intrinsic line
shape as a Gaussian, although we allow for up to two Gaussian
components for emission lines with S/N$>4$ in NGC 4388.  Particularly
in the \brg\ and [Si {\small VI}] lines, the line shapes are often
non-Gaussian and warrant a second component. In $\sim 50-60\%$ of
cases within the central $2 \times 2\arcsec$ the S/N is sufficient to
warrant a two-component fit. We calculate the non-parametric FWHM from
the sum of Gaussian components, and from that value a corresponding
line dispersion.  We also experimented with an empirically determined
line-broadening function by fitting a Gauss-Hermite model to the
strong sky emission lines, which caused only very minor changes in
our fits. We derive errors on all the fitted parameters by fitting 100
artificial spectra that are created as a sum of the original spectrum
and a Gaussian random noise array generated from the extracted errors.

\section{NGC 4388}
\label{sec:s4388}

NGC 4388 is an SBb galaxy at a distance of 19 Mpc \citep{kuoetal2011},
corresponding to a physical resolution of $\sim 50$ pc given our $\sim
0\farcs5$ seeing.  The galaxy magnitude is $M_r = -20.1$ mag, with a
B/T$\sim 0.5$ based on fits to the SDSS data \citep{greeneetal2010}.
We show an HST/WFC3 image combining F336W (broad $U$), F435W (broad
$B$), and F814W (broad $I$) in Figure \ref{fig:image}. With a
photometric P.A. of 90\degr, the galaxy is very nearly edge-on, which
is not true for the majority of the megamaser disk galaxies
\citep{braatzetal1997,greenhilletal2009}.  The inclination determined
from the large-scale kinematics is 78\degr\
\citep[][]{veilleuxetal1999}, and the systemic velocity from the same
study is $v_{\rm sys} = 2525 \pm 25$~\kms.  The gas kinematics show an
isophotal twist on arcmin scales, that \citet[][]{veilleuxetal1999}
successfully model as a bar within the inner 1.5 kpc, with a P.A. of
100\degr\ on the sky. The bar is also apparent in the photometry as a
boxy bulge.  

NGC 4388 is well-known as a galaxy that is falling into
the Virgo cluster \citep{yasudaetal1997}. Gas is being stripped from
the galaxy, as evidenced by an \ion{H}{1} and ionized gas tail
\citep{fordetal1971,phillipsmalin1982,pogge1988,corbinetal1988,
  petitjeandurret1993, veilleuxetal1999b,stoklasovaetal2009}. There is
even a detection of the photoionized gas in the soft X-rays
\citep{iwasawaetal2003}.

We are interested in the center of NGC 4388, in particular because of
the megamaser disk on sub-pc scales \citep{braatzetal2004}.  Based on
the megamaser rotation curve, the BH mass is found to be $8.4 \pm 0.9
\times 10^6$~\msun\ \citep{kuoetal2011}.  Combined with the bolometric
luminosity estimate from \citet{vasudevanetal2013}, the Eddington
ratio is \lledd$\approx 10\%$, which is in very rough agreement with
our estimate from the \oiii\ luminosity \citep{greeneetal2010}.  Like
most megamaser disk galaxies \citep{greenhilletal2008}, the active
galactic nucleus (AGN) in NGC 4388 is Compton thick and quite bright
in hard X-rays
\citep[e.g.,][]{hansonetal1990,takanokoyama1991,iwasawaetal1997,
  forsteretal1999,federovaetal2011}. NGC 4388 has also been seen to
change state in the X-ray \citep{elvisetal2004}, switching from being
Compton thick to Compton thin. Weak broad \halpha\ has also been
reported at the galaxy center
\citep{filippenkosargent1985,hfs1997broad}.

\begin{figure*}
\hbox{
\hskip 15mm
\includegraphics[width=0.8\textwidth]{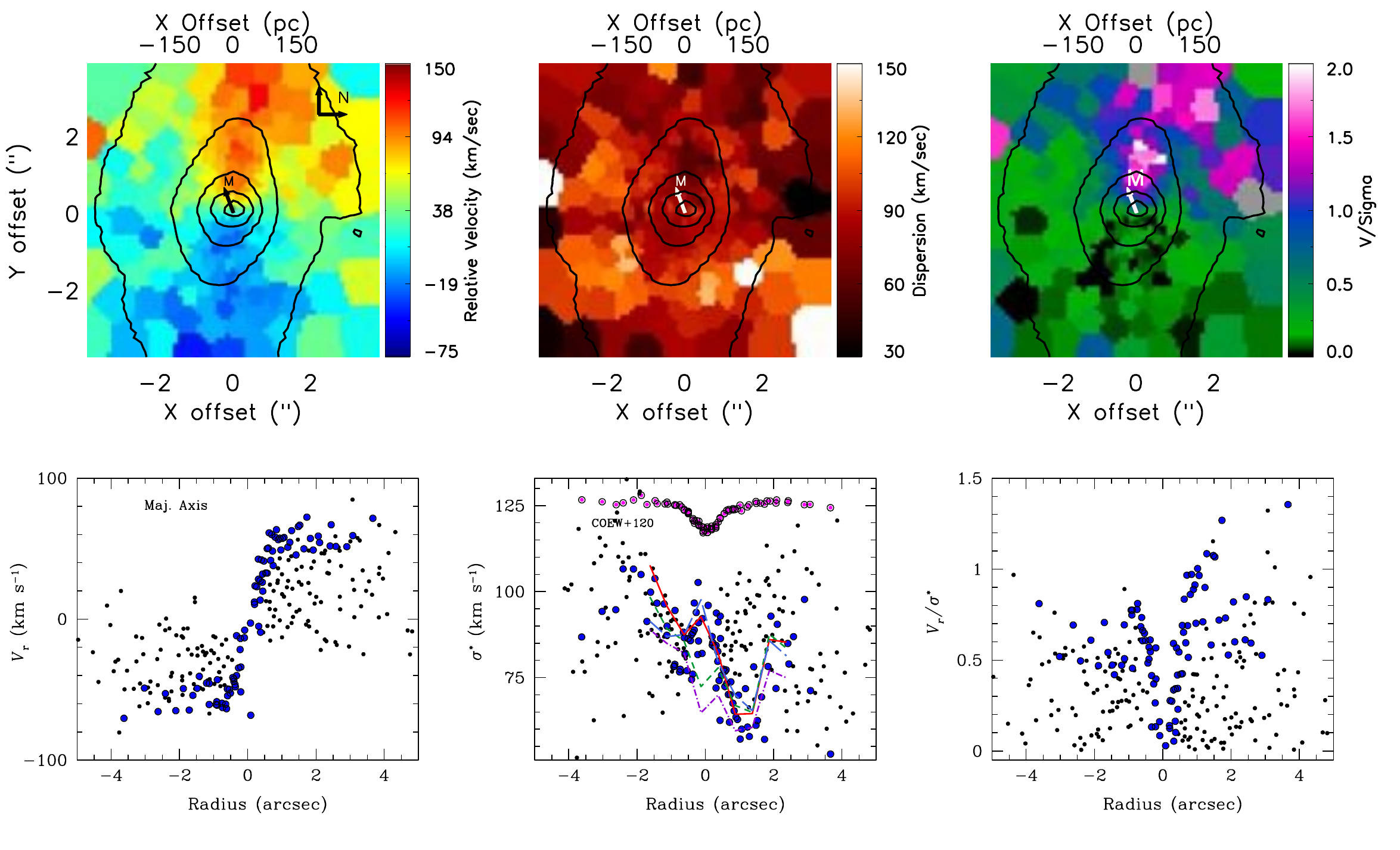}
}
\figcaption{{\it Top}: Stellar rotation (left), velocity dispersion (center), and $V/\sigma$ 
(right) fields as derived from the Voronoi-binned data using pPXF.  The orientation 
has North left and East up, while the sub-pc megamaser disk orientation
is indicated.
{\it Bottom}: Rotation (left), velocity dispersion (center), and $V/\sigma$ 
(right) as a function 
of radius at all position angles (black) and along the major 
axis (blue).  For illustration (center), we also 
show the CO EW along the major axis slice in magenta symbols, scaled up for 
plotting purposes.  Particularly note the asymmetric fall in 
dispersion seen only to one side of the galaxy.  To investigate whether this 
drop in \sigmastar\ is real, or an artifact of the fitting, we refit with 
a polynomial degree of two (red solid line), only K5III templates (green 
dashed) and with Wallace templates using only giant stars (long-dashed 
magenta).  Although the dispersion in the center is uncertain due to AGN 
contamination, the $\sigma$-drop persists in all of these fits.
\label{fig:sigdispfield}
}
\end{figure*}

\subsection{Stellar Kinematics}
\label{sec:stellarkin4388}

We first examine stellar rotation and dispersion.  We clearly see
rotation within our $\sim 300$ pc field of view that aligns with the
disk on larger scales. Following \citet{cappellarietal2009}, we first
use pPXF to fit the average spectrum over all data, excluding a
0\farcs5 radius around the AGN (Figure \ref{fig:specn4388}).  In the
fiducial fits, we do not fit higher-order Gauss-Hermite moments.  We
will examine $h3$ and $h4$ in \S \ref{sec:stelrotcurve} below.  All of our
default fits employ the SINFONI templates.  From this high S/N
spectrum we can derive the best-fit spectral mix, which may be
degenerate in lower S/N individual spectra.  Based on our fit to this
average spectrum, we find a dispersion of $105 \pm 7$~\kms, which
compares well with our previous measurement of $107 \pm 7$~\kms\
\citep{hoetal2009,greeneetal2010}.  The best-fit template is composed
of 50\% K4/5III, 38\% K7III, and 12\% M5 II/III stars.

We then fit spatially resolved spectra across the cube, using Voronoi
binning to ensure comparable S/N in all bins. We both allow the mix of
templates to be a free parameter, and use the best-fit mixture
from the high S/N fit to the entire cube. The latter fit has the
advantage that by decreasing the number of free parameters, 
we increase the fitting stability in the AGN-dominated region
\citep{cappellarietal2009}.  However, we find excellent agreement
between the two measurements, with the scatter in their fractional
difference being only 9\%.  Thus, we will present results with the
stellar templates allowed to vary.

The fitted rotation and dispersion fields are shown in Figure
\ref{fig:sigdispfield}. Our best-fit systemic velocity is $2530$~\kms,
consistent with the measurement of \citet{veilleuxetal1999}.  
We do find a systematic difference of $\sim 20$~\kms\ between the
absolute velocity derived when we use the SINFONI stellar templates
and that from the Wallace et al.\ templates.  We are not sure of the
origin of this difference; it is possible that the SINFONI template
stars, while all at rest relative to each other, have not been shifted
to a rest velocity.  We thus quote an uncertainty of $2530 \pm
20$~\kms\ in the systemic velocity.

\subsubsection{Stellar Rotation Curve}
\label{sec:stelrotcurve}

To derive the rotation curve we use the {\it kinemetry} formalism of
\citet{krajnovicetal2006}, which is similar to ellipse
fitting of photometry, but operates on both even and odd moments of
the kinematic field.  The kinematics are modeled using a sixth order
expansion along ellipses with the position angle and ellipticity 
as free parameters.  We show the resulting rotation
curve in Figure \ref{fig:stelrotcurve}, along with the best-fit
ellipse PA, ellipticity $\epsilon$, and the amplitudes of the higher-order
terms that encapsulate deviations from simple circular rotation.

We clearly measure rotation in the stellar kinematics, with a maximum
amplitude of $V_{\rm max} \approx 60$~\kms.  The stellar rotation axis
of 90\degr\ is aligned with the major axis of the kpc-scale disk.  On
larger scales, \citet{stoklasovaetal2009} measure a stellar rotation of
60-80~\kms\ out to $\sim 4$\arcsec\ ($\sim 300$ pc), consistent with 
ours, which then appears to fall slightly at larger radius.

With \sigmastar$\approx 110$~\kms\ beyond 1\arcsec (100 pc), we find
a $V/\sigma \approx 0.6$.  Overall, the stars are
dispersion-dominated, but we see signs of kinematic components apart
from just a bulge.  Firstly, for a $V/\sigma \approx 0.6$, we would
expect an ellipticity of $\epsilon < 0.3$ for an isotropic oblate
rotator \citep{binney1978}, rather than the $\epsilon \approx 0.5-0.6$
that we observe in the isophotes (Figure \ref{fig:image}).  Secondly,
within a radius of $\sim 1\farcs5$ the dispersion field actually drops
(although the decrement is more pronounced on the East side of the disk). 
On the same scale, we see a jump in
velocity, PA, and ellipticity likely signaling a transition from an
inner disk to more bulge-dominated kinematics.  Thirdly, we also
detect a disk-like structure in the inner 100 pc in the \hst/WFC3
F160W image (Figure \ref{fig:image}).  The nuclear disk appears to
have PA$\, \approx 75$\degr, misaligned with the large-scale disk, and a size
very similar to the $\sigma-$drop region.  We detect this same disk
component in more detailed two-dimensional fitting of the
\hst+ground-based data (R. L{\"a}sker et al.\ in preparation).

The most likely explanation of these observations is that there is a
disk within $2$\arcsec\ (180 pc) embedded in the larger-scale
bulge/bar. Only on unextincted sight lines are the kinematics
dominated by the disk component, while on heavily extincted sight
lines we are seeing the kinematics in the bulge/bar.  The falling
rotation curve outside 400 pc may also reflect the increasing
dominance of the bulge/bar on larger scales. As additional
confirmation of this picture, we refit the stellar kinematics and
measure the higher-order Gauss-Hermite moments $h3$ and $h4$
\citep[e.g.,][]{vandermarelfranx1993}.  Although individual $h3$
measurements have only S/N $\approx 0.5-2$, we do find the classic
anti-correlation between $h3$ and $V/\sigma$ that points to the
superposition of a bulge and disk, as seen both in simulations
\citep[][]{hoffmanetal2009} and data
\citep[e.g.,][]{benderetal1994,krajnovicetal2008,seth2010}.  We also
find two additional hints that the $K-$band extinction is higher to
the West.  Firstly, we find a significantly redder continuum slope to
the West side of the disk, corresponding to a differential $A_V
\approx 2$ mag assuming a \citet{cardellietal1989} reddening law.
Secondly, we clearly see a dip in the \hmol\ emission on the same
scale on the West side of the disk (see \S \ref{sec:gasflux}).

Next we will ensure that the
$\sigma-$drop is real and not an artifact of our fitting procedure.

\includegraphics[width=0.45\textwidth]{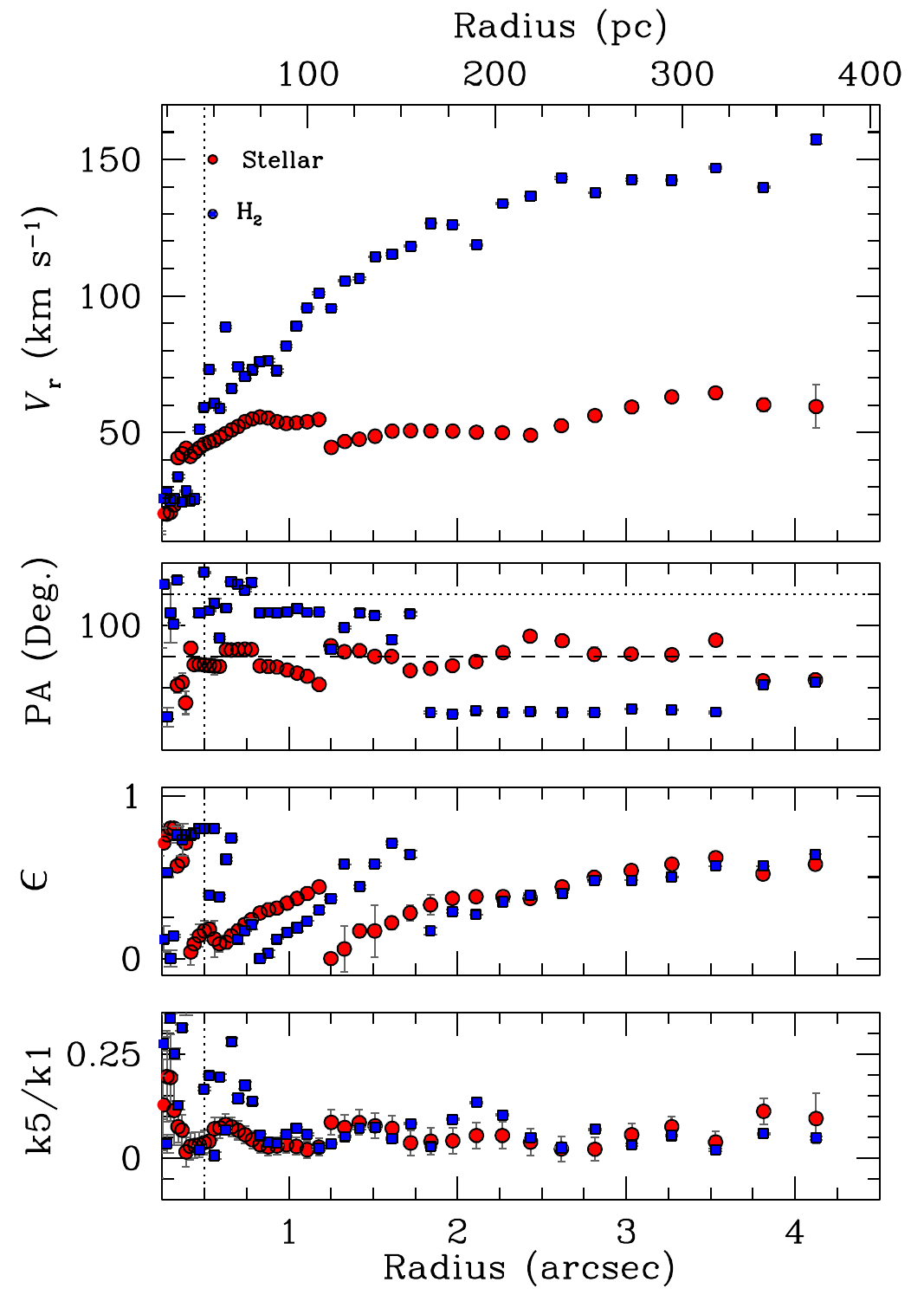}
\figcaption{Properties of the velocity maps as derived from kinemetry
  \citep{krajnovicetal2006} from both the stars (red circles) and and
  the \hmol$~2.12 ~\micron$ gas (blue squares).  We show, in order from
  top to bottom, the rotation curve, the position angle (measured East
  of North), the ellipticity $\epsilon$, and the
  higher-order deviations from a circular velocity pattern.  Our kinematic
  PA is consistent with previous measurements.  For reference, we also 
  show the PA of the large-scale disk (90\degr) and of the megamaser 
  disk (107\degr). We see that most
  of the interesting deviations occur within the inner $\sim 2$\arcsec\
  (200 pc). The gas is considerably colder than the stars, with a
  rotation amplitude of 150 vs 60 \kms\ respectively.
\label{fig:stelrotcurve}
}

\subsubsection{Stellar Dispersion Drop}
\label{sec:sigdrop}

The AGN is the primary source of uncertainty in our modeling of
\sigmastar\ in the galaxy center.  Contamination from the non-thermal
central source dilutes the stellar signal, and is degenerate with
\sigmastar\ \citep[e.g.,][]{greeneho2006sig}.  First, we look at the
S/N of the dispersion measurements as a function of position. We find
that the S/N is always $>5$ except within $0\farcs2$ of the AGN.  This
first test suggests that our measurements at $\sim 1$\arcsec\ (100 pc)
should not be dominated by the AGN.  Next, we re-run pPXF, but instead
of fitting the continuum with the default polynomial of order 4, we
only allow a first or second-order polynomial (Figure
\ref{fig:sigdispfield}).  While it is clear that, particularly in the
case of the first-order polynomial, we achieve very poor fits to the
nuclear region ($\lesssim 0\farcs2$), we still find significant
evidence for an asymmetric $\sigma-$drop.  We then return to the
Wallace templates (rather than our default SINFONI templates) and try
restricting the template set to K dwarfs.  We find the same result.
Finally, we examine our fits with the template fixed to the mixture
derived from the high S/N fit described in \S
\ref{sec:stellarkin4388}. In all of these cases, although the
dispersion profile within 0\farcs5 is not well determined, we recover
the $\sigma-$drop on 1\arcsec\ (100 pc) scales.  We therefore conclude
that the observed drop is real and that the asymmetry results
from patchy reddening.  Although in general the North side of the disk
is more heavily extincted (\S \ref{sec:nlr}), there are also dust
lanes extending to the South, which are apparently preferentially
affecting the West side of the 100-pc--scale disk.

\begin{figure*}
\hbox{
\hskip 15mm
\includegraphics[width=0.8\textwidth]{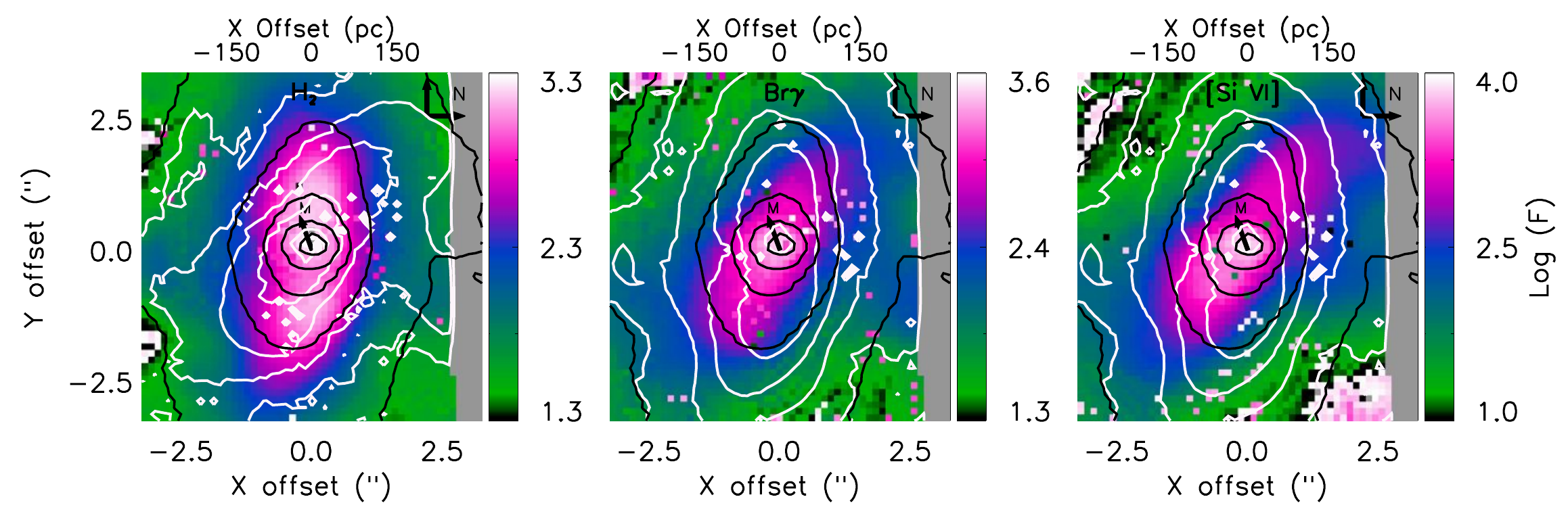}
}
\figcaption{
We show the integrated flux in \hmol$~2.12 ~\micron$ 
(left), Br$\gamma$ (middle), and [Si {\tiny VI}] (right) for the SINFONI 
data cube. In black contours we show the galaxy continuum 
(same for all panels).  In white, for comparison 
we show the Br$\gamma$ flux (left), and the \hmol\ flux (middle and 
right). Since the data are not flux calibrated, the flux scale is 
arbitrary.  Note first of all the very different flux distributions for the 
molecular hydrogen and the other two lines.  Specifically, the \hmol\ 
is oriented E-W along the nuclear disk seen in the \hst\ imaging, 
while the other lines are at an angle nearly perpendicular to the maser disk, that 
matches the inner jet.  It is also interesting to note that the \hmol\ 
distribution is quite asymmetric.  Very likely the same extinction that 
hides the W side of the nuclear stellar disk and causes the asymmetric 
$\sigma-$drop is also causing this asymmetric distribution in the 
molecular gas.  In contrast the [Si {\tiny VI}] and Br $\gamma$ are 
symmetric in the center, and if 
anything are brighter towards the West.
\label{fig:emissionflux}
}
\end{figure*}

A falling dispersion towards the center of megamaser disk galaxies
is not unexpected.  These so-called ``$\sigma$-drops'' are
common both in spiral
\citep[e.g,][]{bottema1993,emsellemetal2001,falconbarrosoetal2006,
  peletieretal2007,riffeletal2011}, and in early-type galaxies
\citep{pinkneyetal2003,emsellemetal2004, lyubenovaetal2008}, in the latter often
associated with nuclear star clusters \citep{lyubenovaetal2013}.  The
typical scale is $\sim 200$ pc, and the most likely explanation for
the falling dispersion is that the galaxy harbors a gas disk that is
currently or was recently forming stars.  These bright, dynamically
cold stars dominate the central dispersion and cause \sigmastar\ to
fall.  In fact, NGC 2273, one of the nearer megamaser disk galaxies,
has a dramatic $\sigma-$drop that coincides with a known disk in the
inner $\sim 500$ pc
\citep{erwinsparke2003,barbosaetal2006}. \citet{hicksetal2013} suggest
that low central dispersions are more common in active galaxies.

If the $\sigma-$drop is caused by a young stellar disk, then we might
also expect to see younger stars coinciding with the lower stellar
velocity dispersion.  We create a median spectrum over the region with
\sigmastar$<80$~\kms, and rerun pPXF to investigate the mixture of
stellar populations using the Wallace templates, which cover a wider
range of spectral type. While there is a slight preference for more
supergiant light in the region of the $\sigma-$drop, we do not find a
significant difference in best-fit spectral type as a function of
region.  The CO EW profile is also quite symmetric and has reached its
asymptotic values already in the $\sigma-$drop region (Figure
\ref{fig:sigdispfield}).  With a larger spiral sample,
\citet{peletieretal2007} also find similar stellar population ages
inside and outside of the $\sigma-$drop regions.

\subsubsection{Stellar Kinematics Summary}

We propose that the observed stellar kinematics are the superposition
of three distinct components.  On kpc-scales, the galaxy disk has
PA$=90$\degr.  Within our 300 pc aperture, the kinematics are
dominated by the hotter bulge/bar \citep{veilleuxetal1999}.  Based on
the highest \sigmastar\ measurements within the cube, presumably
occurring where dust obscuration is highest, we measure
\sigmastar$\approx 125$~\kms\ in the bulge/bar.  Then, within the
inner 100 pc, we see evidence in both the kinematics and the NIR
isophotes for a nuclear disk component at a PA$\approx 75$\degr,
misaligned by $\sim 15$\degr\ from the large-scale disk and with a
$V/\sigma \approx 1$.  Finally, on pc scales, there is a masing disk
at PA$=110$\degr, which is misaligned not only from the kpc-scale
disk, but also by $\sim 35$\degr\ from the nuclear stellar disk
\citep{greeneetal2013}.

\subsection{Gas Fluxes and Kinematics}
\label{sec:gasflux}

With a best-fit stellar rotation field in hand, we turn to the gas
kinematics.  We will examine three transitions: \hmol$~2.12 ~\micron$
1-0S(1), \brg, and the coronal line [\ion{Si}{6}]$~\lambda 1.96
~\micron$.  We note that the [Ca {\small VIII}]$~\lambda 2.32
~\micron$ line, also coronal, has been used to trace the NLR
\citep[e.g.,][]{storchibergmannetal2009}, but is blended with the CO
bandhead.  Thus we choose to focus on the [\ion{Si}{6}] line here.
The two-dimensional velocity and line-ratio fields will provide
information about the origins and excitation mechanisms of each
transition.  The fluxes that we present are the integrated flux from
our one or two Gaussian models as described in \S3.3.  We calculate
the FWHM non-parametrically from the sum of both components, and the
dispersions are simply calculated as FWHM/2.35.  We also measure and
present the skewness of the lines, which encapsulates the asymmetry in
the two-component fits.

\begin{figure*}
\includegraphics[width=0.95\textwidth]{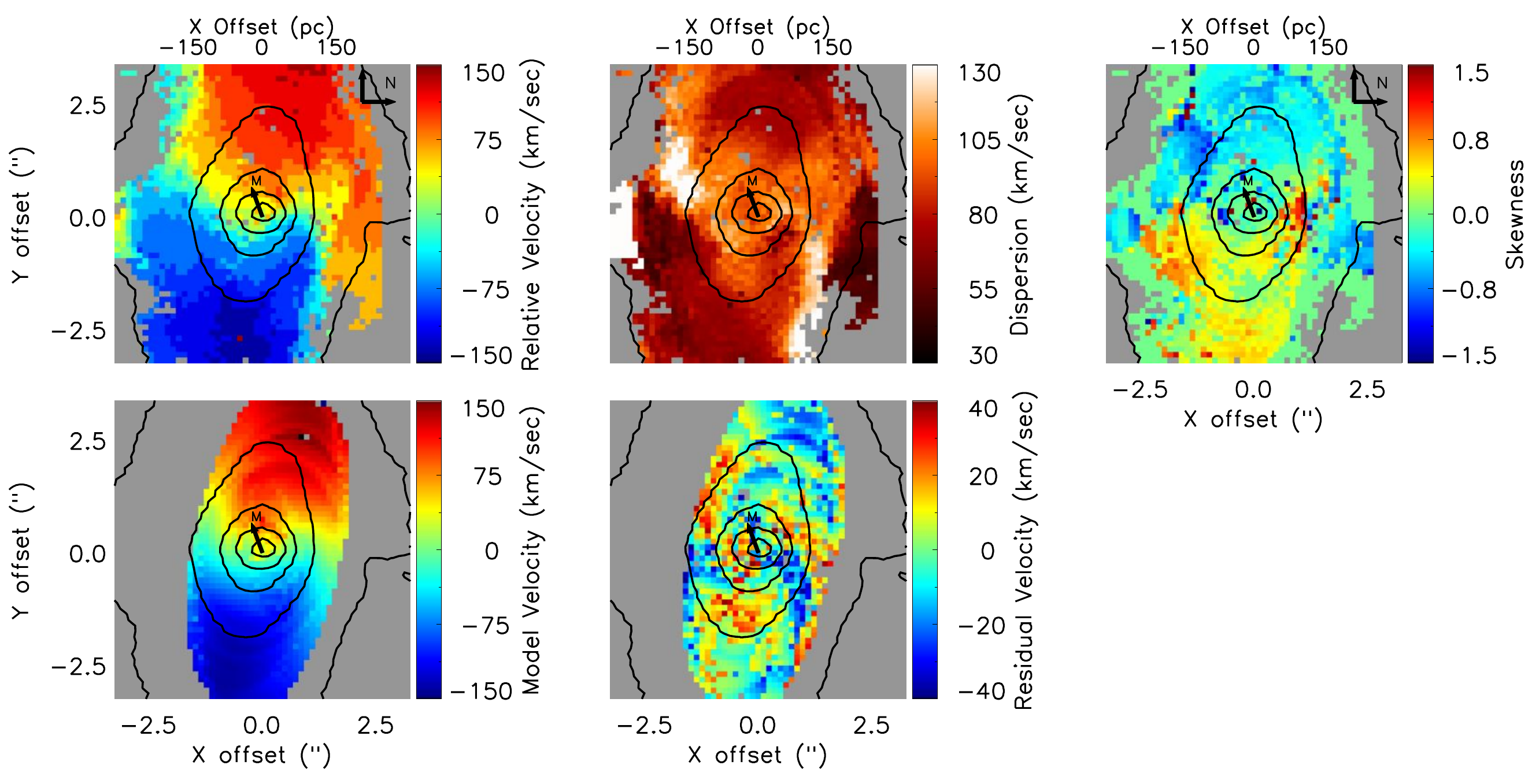}
\figcaption{
In the top row we show the \hmol$~2.12 ~\micron$ velocity (left), dispersion (middle), and skewness (right) fields.  On the bottom row, we show the best-fit circular velocity from a tilted-ring model using {\it kinemetry} (left), and the difference between the velocity field and the model (right).  The contours are continuum flux density, as in Figure \ref{fig:sigdispfield}. The orientation of the cube has North to the right and East up, as shown, and we also indicate the orientation of the megamaser disk on sub-pc scales (107\degr). The rotation field shows a kink in the center, with the inner PA intriguingly close to that of the maser disk. There are corresponding spiral-shaped arms of high dispersion, and a three-armed pattern in the model residuals. All of these may be signatures of a nuclear disk, inflow/outflow out of the plane of the disk, or possibly the inward extension of the large-scale bar.
\label{fig:h2rot}
}
\end{figure*}

In Figure \ref{fig:emissionflux}, we show the distribution of fluxes
in the \hmol, \brg, and [\ion{Si}{6}] emission lines, while in Figures
\ref{fig:h2rot}, \ref{fig:oiiisivi}, and \ref{fig:2dgas}, we summarize
their kinematics.  We see that the overall morphology of the \hmol\
follows the stellar continuum, and is even more flattened than the
stellar distribution, suggesting that the emission arises from the
nuclear disk. Furthermore, close inspection reveals that the \hmol\
emission is asymmetric, being considerably brighter towards the East.
The brighter side of the \hmol\ disk is coincident with the observed
$\sigma-$drop in the stellar distribution; we can directly see the
extinction of the \hmol\ in the region where the $\sigma-$drop is also
extincted. In contrast, the high-ionization lines are oriented at
PA$\approx 30\degr$, aligned with the jet on similar scales, and also
with the \oiii\ emission that traces the narrow-line region (NLR;
Fig. \ref{fig:oiiisivi}).  The \brg\ mostly follows the
high-ionization lines (Fig. \ref{fig:2dgas}). Their light distribution
strongly suggests that their emission arises in front of the disk in
projection, since the [\ion{Si}{6}] is actually brighter towards the
West (similar to the \oiii).  Based partially on these flux
distributions, as well as the kinematics and line ratios presented
below, we will suggest that the \hmol\ mainly traces star formation in
the nuclear disk, while \brg\ and [\ion{Si}{6}] trace the NLR
\citep[e.g.,][]{rodriguezardilaetal2004,
  storchibergmannetal2009,riffelstorchibergmann2011,
  mazzalayetal2013}.

\subsubsection{Molecular Hydrogen Kinematics}

We focus first on the \hmol$~2.12~\micron$ kinematics (Figure
\ref{fig:h2rot}).  As with the stellar kinematics above, we fit the
\hmol\ velocity field using kinemetry.  The best-fit rotation curve,
PA, and eccentricity are shown in blue in Figure
\ref{fig:stelrotcurve}.  With V/$\sigma \approx 1.5$, the gas is
considerably colder than the stars, but we are not seeing pure
rotation. The rotation field shows an $S$-shaped kink.  The twist is
apparent as a discontinuity in the \hmol\ ellipticity $\epsilon$ and
PA in the rotation profiles at $1\farcs5$ (150 pc), a similar radius
to the changing stellar kinematics.  In the skewness map, along
  the high-dispersion kink, we see a tendency for red asymmetry, which
  is plausibly attributed to the superposition of two velocity
  components. Interior to the twist, the kinematic PA of the gas on
$<150$ pc scales appears to align with the megamaser disk.  Thus, it
is possible that whatever mechanism causes this kink is responsible
for supplying material on pc scales. 

On the same scales, we find a two-armed spiral in the dispersion map,
with an inner PA $\approx 30\degr$. We find a corresponding three-arm
spiral pattern in the residual map by subtracting the best-fit
rotation model from the observed velocity field. To
interpret these trends, we recall that any $m = 2$ mode perturbation
can cause $m^{\prime} = 2$ deviations in the even velocity moments,
  including density and dispersion, and due to geometry $m^{\prime} =
  3$ (and $m^{\prime} = 1$) deviations in the odd velocity moments
  \citep[e.g.,][]{canzian1993}. We propose a few mechanisms that 
may explain these kinematic properties.

First, in the study by \citet{vandevenfathi2010} of the inner region
of NGC 1097, similar signatures were explained as the result of a
two-arm nuclear spiral in the disk plane \citep[see
also][]{daviesetal2009}. Given that the nuclear disk in NGC4388 is nearly
edge-on, such a co-planar spiral density wave seems unlikely. The
second possibility is that we are seeing the superposition of the
kpc-scale disk and the 100 pc nuclear disk that we saw in the stellar
kinematics.  However, the gas PA$\approx 110\degr$, while
approximately aligned with the maser disk on sub-pc scales, does not
align with the stellar disk of PA$\approx 75$\degr.  The final
possibility is that the twist we observe here is related to the kink
from the bar observed on kpc scales \citep[e.g.,][]{veilleuxetal1999}.
Only with our upcoming AO observations can we address the $<100$ pc
gas kinematics definitively, and hopefully help determine what is
driving the gas inward towards the AGN.

\hbox{
\hskip 6mm
\includegraphics[width=0.4\textwidth]{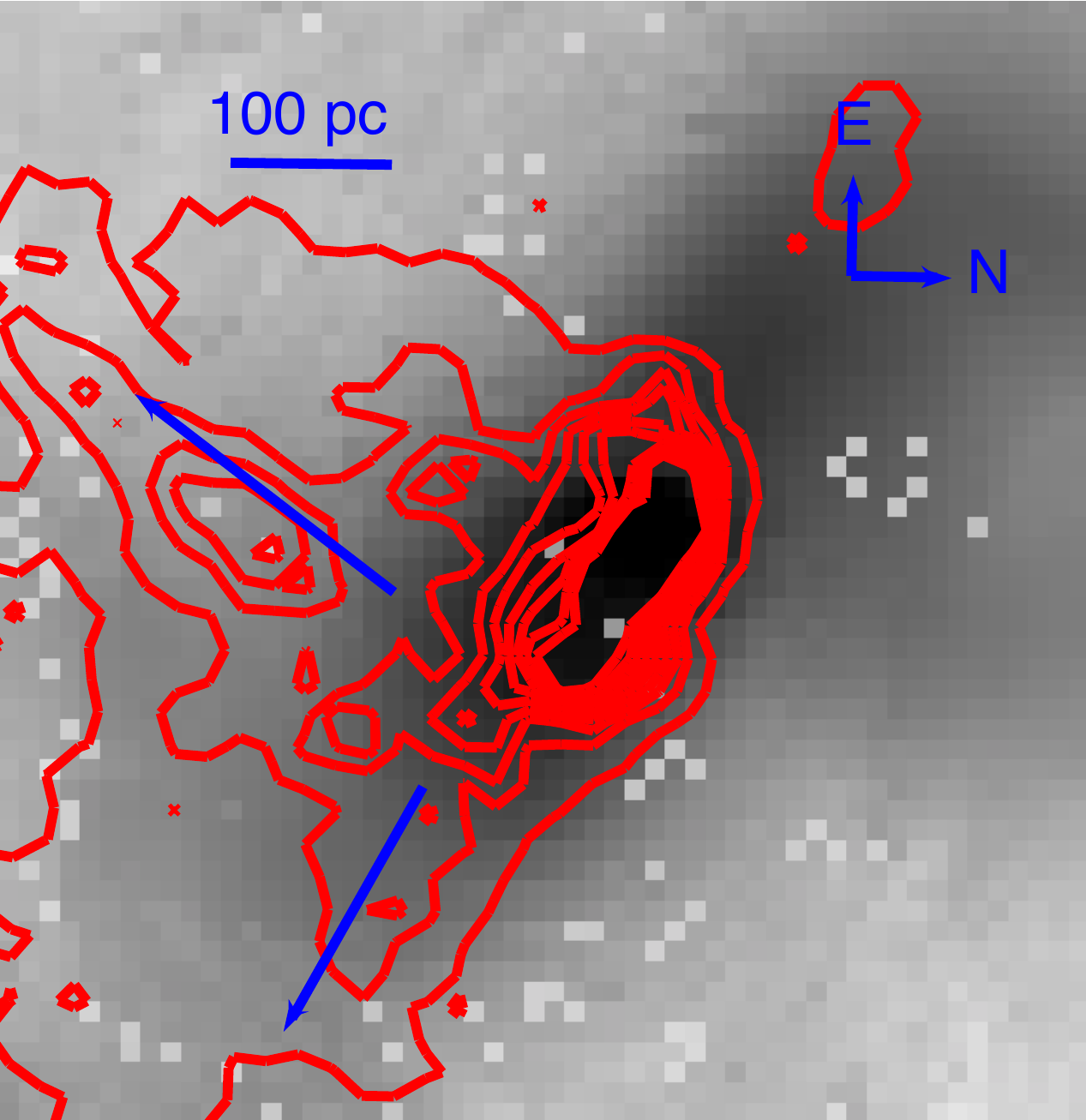}
}
\figcaption{We show the correspondence between the optical and NIR 
  high-ionization lines. 
  In greyscale we show the [Si {\tiny VI}] emission map over the same
  field-of-view displayed in all other maps above.  We use [Si {\tiny VI}] 
  to represent the purest and highest S/N view of the NLR as traced 
  by emission lines within our cube. The
  narrow-band [O {\tiny III}]$~\lambda 5007$~\AA\ image from \hst\
  is overlaid as red contours.  The scale bar indicates $1\arcsec$ ($\sim
  100$pc). The blue arrows schematically indicate the edges of the
  ionization cone that we see hints of in the [Si {\tiny VI}] and
  Br$\gamma$ maps.  The radio emission as imaged by
  \citet{falckeetal1998} follows the ionized gas closely on these
  scales, and then opens into a plume to the North-East (off of this
  image).
\label{fig:oiiisivi}
}
\vskip 5mm
\noindent

\subsubsection{Molecular Hydrogen Excitation}
\label{sec:h2excitation}

\begin{figure*}
\includegraphics[width=0.95\textwidth]{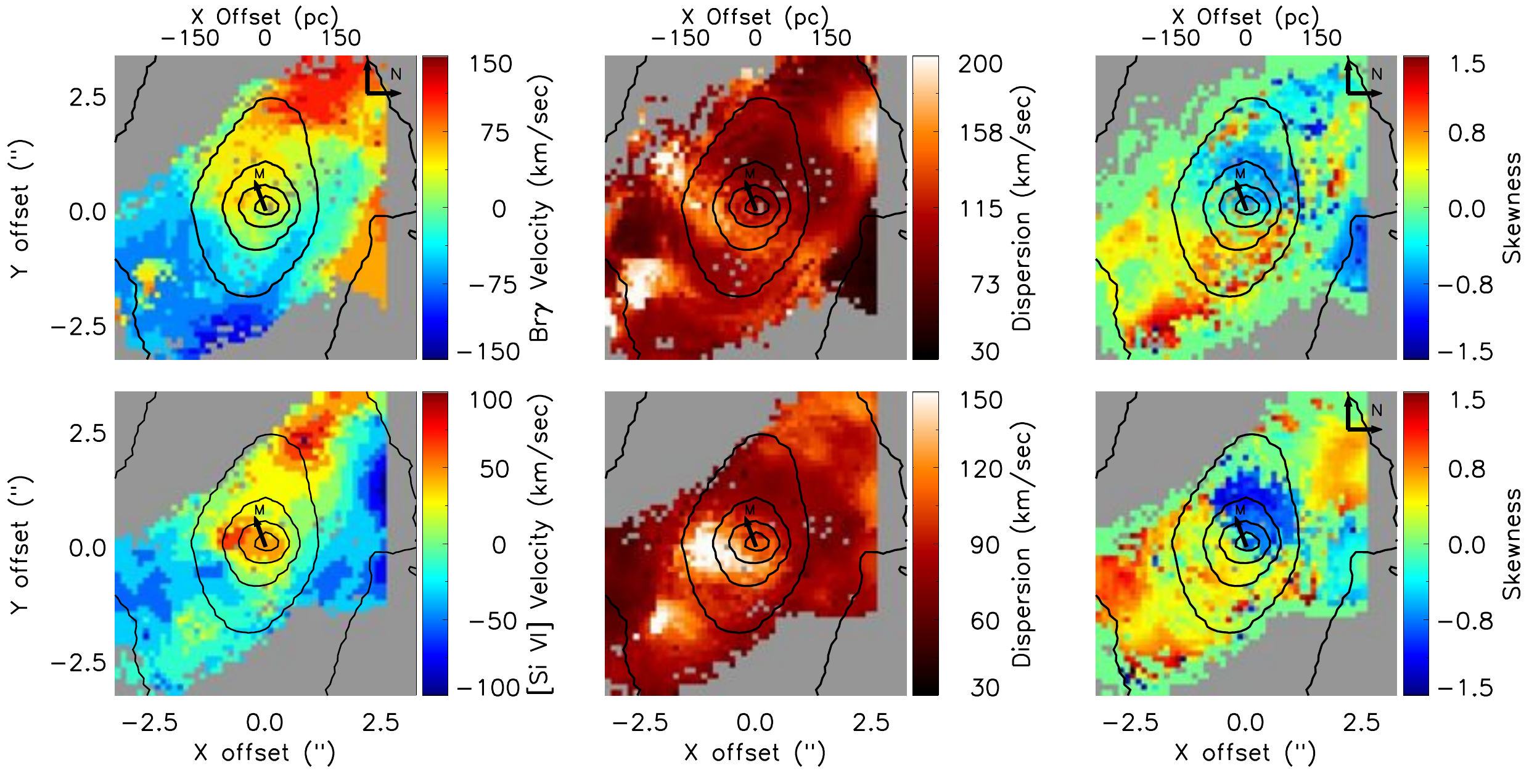}
\figcaption{Two-dimensional gas velocity (left), dispersion (middle), 
  and skewness (right) for \brg\
  (top) and [Si {\tiny VI}] (bottom), contours show galaxy continuum
  as above.  In all cases, the images are aligned with East up and
  North to the right, to match the data cube itself.  The molecular
  hydrogen (Figure \ref{fig:h2rot}) is aligned with the kpc-scale disk
  of the galaxy, while these transitions are more closely aligned with
  the pc-scale radio jet.
\label{fig:2dgas}
}
\end{figure*}

\begin{figure*}
\includegraphics[width=0.95\textwidth]{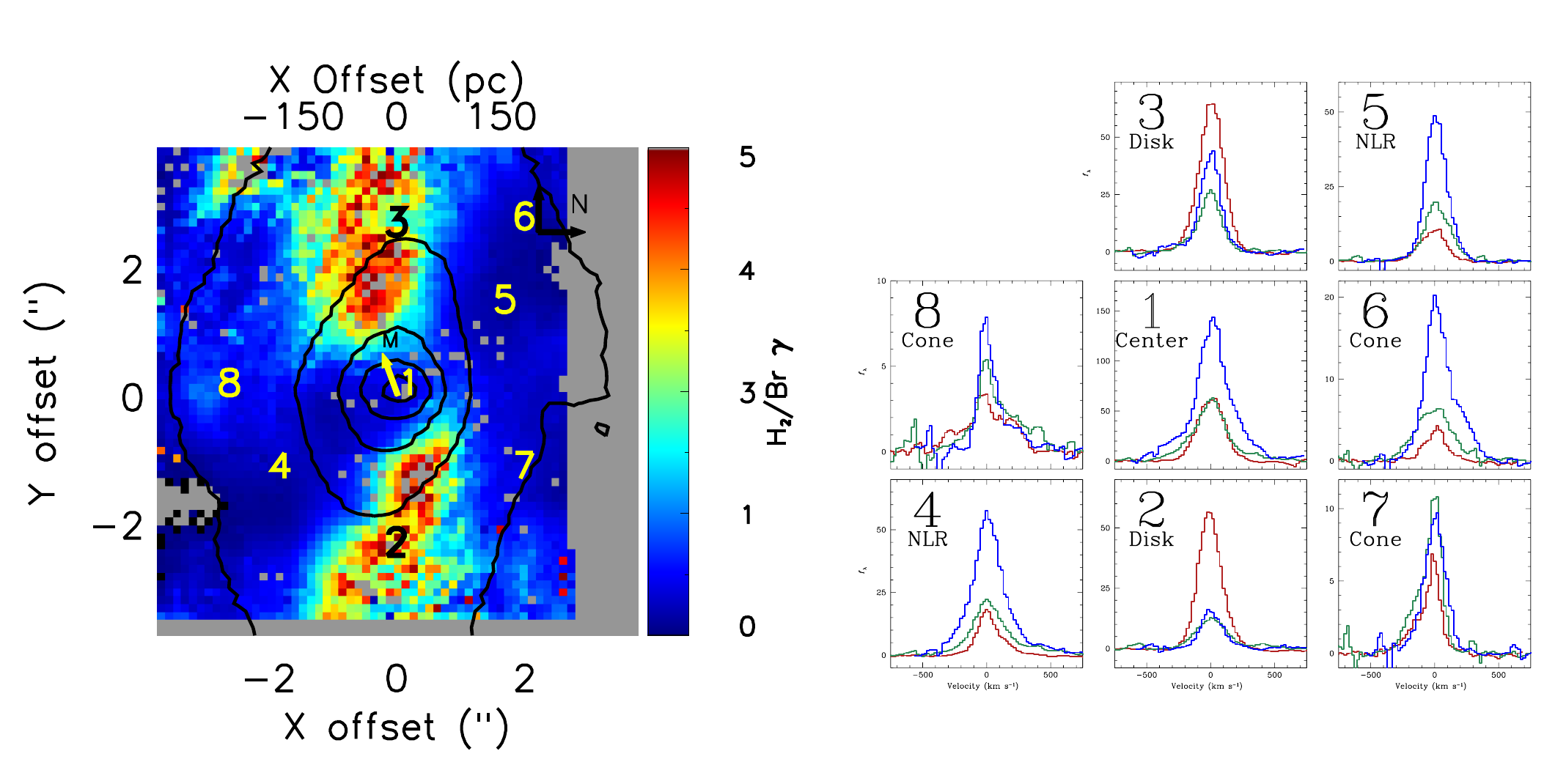}
\figcaption{{\it Left}: The \hmol$~\lambda 2.12 ~\micron$/\brg\ ratio.  The 
values are high in the disk, and quite low along the jet/NLR direction, 
perpendicular to the maser disk. Each number marks a region that 
we have extracted to create higher S/N spectra.
{\it Right}:  Spectra in each region, of
  H$_2$ 1-0S(1) (2.12 ~\micron; red), Br$\gamma$ (green) and [Si {\tiny VI}] (blue).
  The flux normalizations are arbitrary, but the spectra span
  $700$~\kms\ in velocity on either side of systemic.  
\label{fig:h2brg}
}
\end{figure*}

\begin{figure*}
\hbox{
\hskip 15mm
\includegraphics[width=0.8\textwidth]{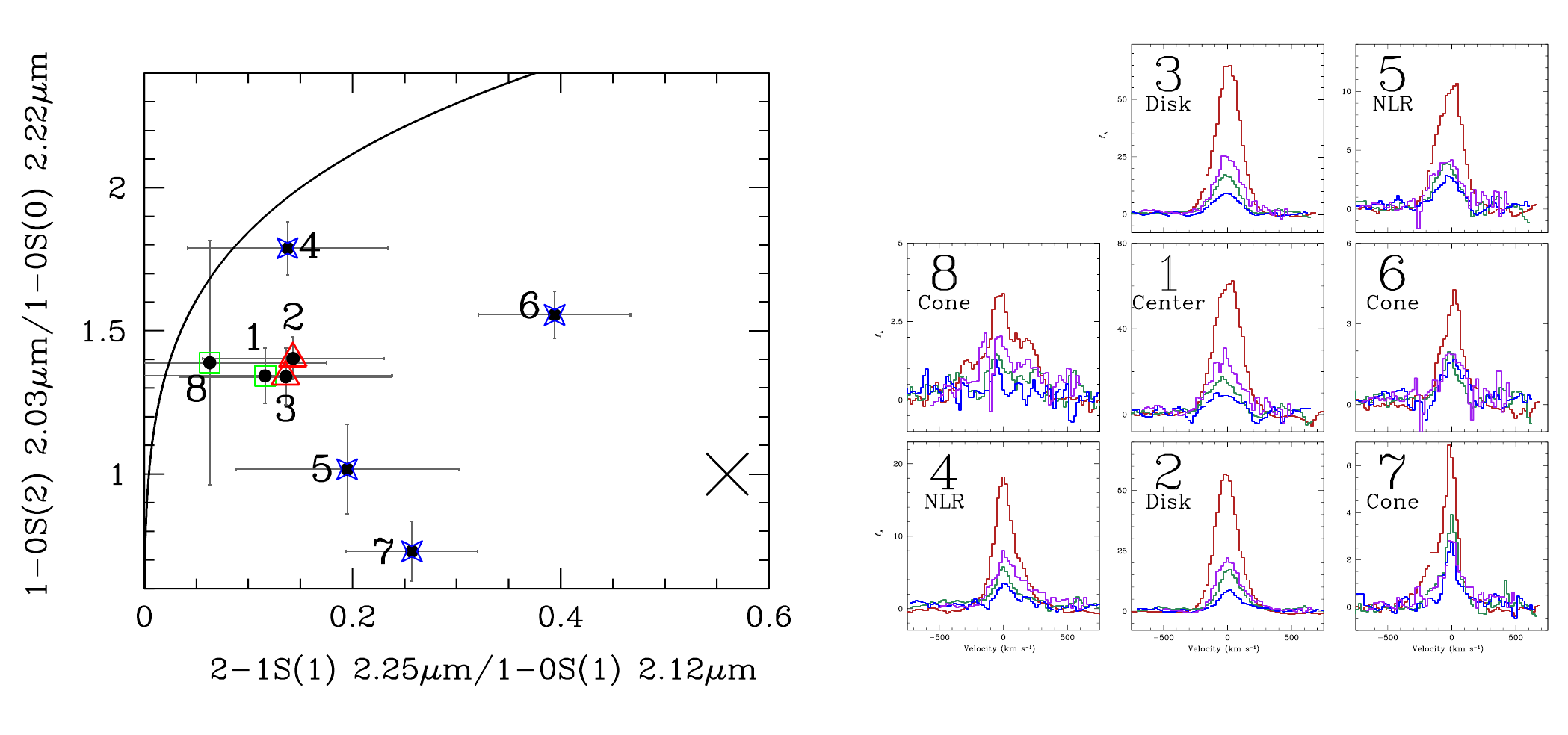}
}
\figcaption{{\it Left}: 
  \hmol\ line ratio diagram, comparing the ratio of \hmol\ 2-1S(1) 
   $2.25 ~\micron$ /1-0S(1) $2.12 ~\micron$ to the ratio of
  \hmol\ 2-0S(2) $2.03 ~\micron$/1-0S(0) $2.22 ~\micron$.  The solid line
  shows the theoretical expectation for pure thermal excitation, 
  while the large black cross schematically 
  shows the region occupied by pure fluorescence models 
  in \citet{blackvandishoeck1987}.  The measurements are made on the
  spectral regions described in the text (\S \ref{sec:h2excitation}) and in
  Table 1, and have been divided into those with low ratios of
  \hmol/Br$\gamma$ ($\leq 0.6$; blue stars), medium ratios between 0.6
  and 2 (green squares), and high ratios greater than 2 (red
  triangle). The first is characteristic of star forming regions, the
  second of Seyfert galaxies, and the final of LINERs.
  {\it Right}: 
  We show the spectra of all four \hmol\ transitions in each 
  region.  The strongest (red) is \hmol\ 1-0S(1) $2.12 ~\micron$, next 
  (purple) is \hmol\ 2-0S(2) $2.03 ~\micron$, then (green) 
  \hmol\ 1-0S(0) $2.22 ~\micron$ and finally weakest (blue) is \hmol\ 
  2-1S(1) $2.25 ~\micron$.  
  The velocity structure of all transitions 
  is very similar in the disk, while there are interesting differences in 
  the NLR and ionization cone regions.  Particularly note that in region 7, 
  the 1-0S(1) line has a blue wing while the 2-1S(1) line has a red wing.
\label{fig:h2rat}
}
\end{figure*}

The \hmol\ shows very regular rotation along the orientation of the
large-scale galaxy disk outside of the inner 1\arcsec\ (100
pc). Additional information is encoded in the ratios of \hmol\ to
\brg\ (Figure \ref{fig:h2brg}), and ratios of different \hmol\
transitions (Figure \ref{fig:h2rat}).  Typically, in the hard ionizing
field of the NLR, the ratio of \hmol\ to \brg\ takes a value $0.6 <
$\hmol/\brg$ < 2$, while it is $\leq 0.6$ in star-forming regions
\citep{larkinetal1998,moorwoodoliva1990,
  rodriguezardilaetal2004,rodriguezardilaetal2005,riffeletal2010}.
Finally, in dense spiral arms, even higher ratios of \hmol/\brg$ > 2$
are observed. \citet{riffeletal2013h2} suggest that the increased
\hmol\ emission arises from the higher gas densities within spiral
arms \citep[see also][]{riffeletal2013}.  NGC 4388 obeys these trends
in general (Figure \ref{fig:h2brg}). Along the disk, the \hmol\ is
strong, and the ratio has values $>2$, as is commonly seen in spiral
arms, while in the galaxy center, the line ratio ranges from $0.6
<$\hmol/\brg$<1$, values common for Seyfert galaxies.

The only surprising region lies off the disk along a $\sim 60$\degr\
angle East of North above and below the galaxy plane, where the \brg\
emission is strongest. Here, 
the \hmol/\brg\ ratio is very low $<0.5$.  Such low ratios are
reportedly associated with star-forming regions
\citep[e.g.,][]{moorwoodoliva1994,larkinetal1998,
  rodriguezardilaetal2004,rodriguezardilaetal2005,riffeletal2013h2}.
However, in this case, the \brg\ and [Si {\small VI}] morphologies
match both the orientation of the radio jet and the NLR traced by 
\oiii$~\lambda 5007$ in the optical.
Thus, we suggest that the low \hmol\ relative to \brg\ in these
regions is instead due to very low densities of the NLR \citep{dorsetal2012}.

We now construct spectra in characteristic regions to examine both the
\hmol\ line ratios and the line kinematics. We select regions of the
map based on the \hmol/\brg\ line ratio and construct high S/N spectra
from the continuum-subtracted cubes.  Region 1 is the center, defined
as all pixels with CO EW$\leq 6$\AA\ that lie within $1\farcs5$ of the
center.  We extract two regions in the disk (H$_2$/\brg$<0.6$; regions
2 \& 3) and two regions in the NLR (H$_2$/\brg$>1$; regions 4 \& 5).
Finally, we sample gas that falls beyond the main galaxy and overlaps
with the \oiii\ ionization cone (region 8), as well as two regions
North of the disk, towards the Northern radio lobe (regions 6 \& 7).
The extracted spectra for the strong transitions (\hmol$~2.12
\micron$, \brg, and [Si {\tiny VI}]) from each region are shown in
Figure \ref{fig:h2brg}, the spectra from different \hmol\ transitions
are shown in Figure \ref{fig:h2rat}, and measured properties from
these spectra are presented in Table 1.  Because we fit the lines
using an instrumental profile, the linewidths are naturally corrected
for instrumental resolution.  Each line is fitted independently.  In
fact, as shown in Figure \ref{fig:h2rat}, the centroids and line
shapes for the different transitions can differ, particularly for
regions in the NLR ionization cone.

For each region, we can ask about the excitation mechanisms of the
\hmol.  There are two ways to excite the warm \hmol: thermal processes
or fluorescence via UV radiation. The thermal processes that can in
principle heat the gas include shocks, X-ray emission, or heating by
UV emission.

The predicted line ratios of various \hmol\ transitions provide diagnostics
of the primary excitation mechanism.  Specifically, if fluorescence
dominates the excitation, then the ratio of \hmol\ 2-1S(1)
2.25$~\micron$/1-0S(1) 2.12$~\micron$ is expected to be $\sim 0.5-0.6$
while the 1-0S(2) 2.03$~\micron$/1-0S(0) 2.22$~\micron$ ratio is
expected to be $\sim 1$ \citep[][]{blackvandishoeck1987}.  We show
this pair of ratios and the expectation for pure thermal emission as
the solid line, while the region dominated by fluorescence is shown
schematically with the large X at the lower right in Figure
\ref{fig:h2rat}.  There is room for some UV fluorescence, but these
line ratios suggest that the bulk of the excitation is thermal.
Similar conclusions have been drawn for larger samples of Seyfert
galaxies \citep{veilleuxetal1997,quillenetal1999,rigopoulouetal2002}.

\vskip 5mm

Our excitation diagram is constructed for the integrated line profile
but, in fact the line ratio is clearly a function of velocity,
particularly for the line emission arising from the NLR ionization
cones.  We do not have sufficient S/N to construct line ratios for the
line core and line wings separately, but these different line shapes
are pointing to different dominant excitation mechanisms in the line
wings.  Higher S/N is needed to quantitatively examine
velocity-dependent line ratios.

We are still left with the question of which thermal process dominates
the excitation.  Previous work has drawn various conclusions, with
\citet[][]{veilleuxetal1997} for instance favoring shocks in outflows,
\citet{quillenetal1999} favoring a mix of processes including emission
from photo-dissociation regions, \citet{storchibergmannetal2009}
favoring X-ray heating or shocks, and so on.  Recent papers have
favored X-ray heating in the vicinity of an AGN
\citep[e.g.,][]{hicksetal2009}.  However, in our case, the fact that
there is little \hmol\ coming the NLR while the bulk of the \hmol\
emission on these scales follows the morphology of the starlight
(Figure \ref{fig:emissionflux}) strongly suggests that the \hmol\ is
predominantly excited by stellar processes (e.g., UV radiation and/or
shocks from supernovas) rather than the AGN.

\subsubsection{The Narrow-line Region}
\label{sec:nlr}

Bridging the gap between the accretion flow on $<$pc scales and the
large-scale ionized gas outflow, we can examine 
the ionized gas emission on 100 pc scales.  In this region, the
radio emission is complex, extends predominantly North-South
(perpendicular to the disk) and contains two unresolved peaks and a
more diffuse ``bubble'' to the north
\citep{stoneetal1988,hummelsaikia1991}. \citet{falckeetal1998} show
that the Northern radio source is coincident with the optical nucleus,
while the Southern radio knot is coincident with a bright region of
ionized gas seen in \oiii, strongly suggesting a jet-ISM interaction.  
High-ionization gas is present to the South of the nucleus in our
data cubes as well.  Previous near-infrared spectra
\citep[e.g.,][]{wingeetal2000, knopetal2001,imanishi2003} show
extended \hmol\ coincident with the galaxy disk, and high-ionization
gas emission from \brg\ and [\ion{Si}{6}] coincident with the nucleus
and ionization cone.  With our high spatial resolution and
three-dimensional coverage, we can robustly disentangle the disk
and NLR kinematics.

The \brg\ and \sivi\ emission lines show similar spatial distributions
that parallel the NLR emission seen in \oiii\ (Figures
\ref{fig:emissionflux} and \ref{fig:oiiisivi}).  Their emission is
distributed at an angle of PA$\approx 30\degr$, tracing the general
orientation of the radio jet seen on similar scales
\citep{falckeetal1998}. The ionization cone seen in \oiii\ continues
to the South-West beyond the SINFONI field-of-view
\citep{pogge1988,veilleuxetal1999,stoklasovaetal2009}, and is filled
with radio plasma \citep{falckeetal1998}. To the North, beyond our
field of view, the jet opens into another wide angle fan.  It is
likely that the Southern ionization cone sits in front of the galaxy
disk, while the Northern one (which we can see in this map and in the
radio but not in the optical) is extincted by the galaxy disk.

\vbox{ 
\vskip +4mm
\hskip -3mm
\includegraphics[width=0.5\textwidth]{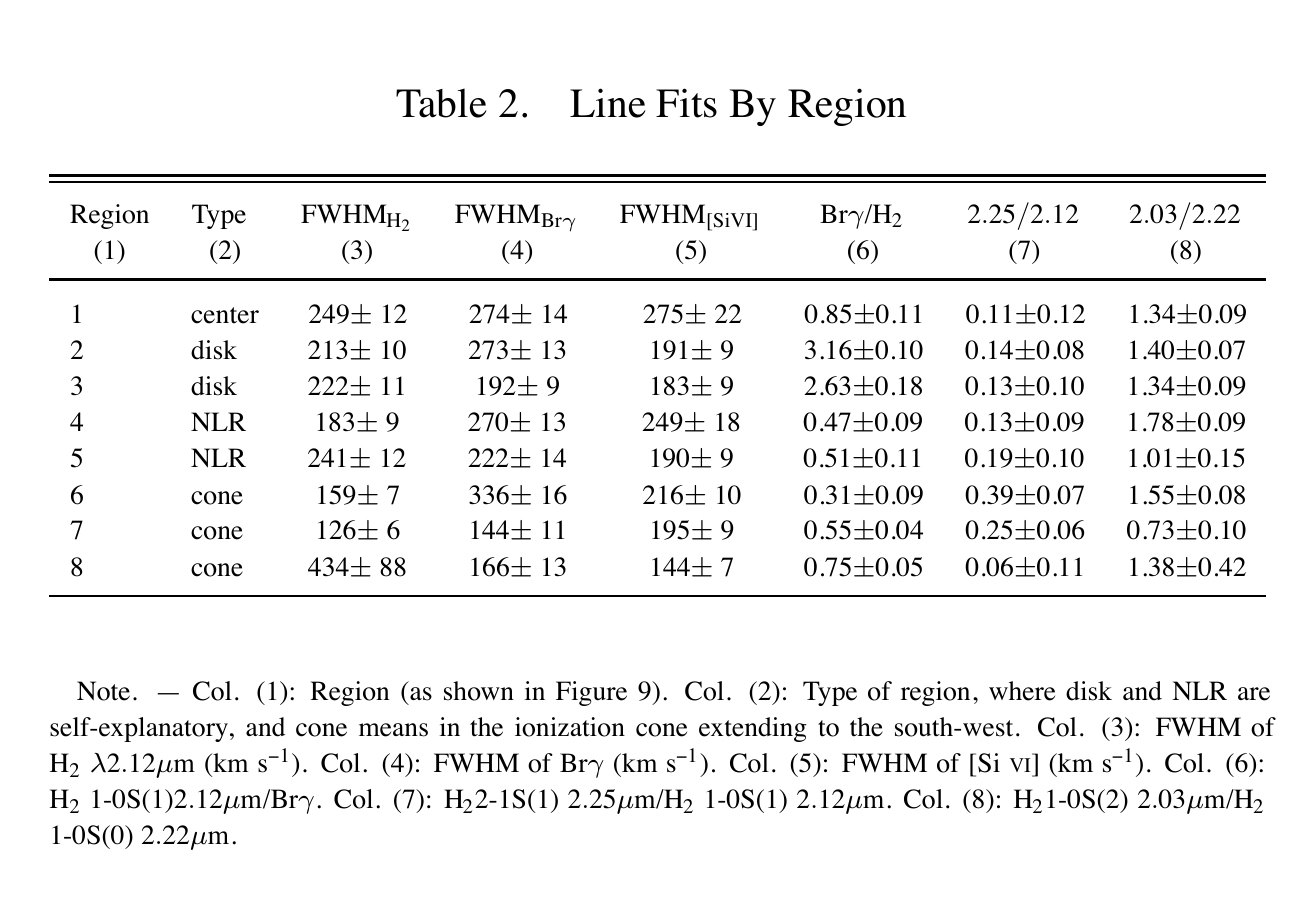}
}
\label{tab:region}
\vskip 4mm

In Figure \ref{fig:2dgas}, we show the velocity and FWHM fields for
\brg\ and \sivi. Both the morphology and the kinematics differ from
that seen in the \hmol.  Specifically, we do not see coherent rotation
in either transition.  The skewness plots are also interesting. In
  both transitions we see redshifted gas towards the South-West and
  blue-shifted gas towards the North-East.  Our interpretation is that
  we are seeing bulk motion of the gas along the jet, which apparently
  points towards us to the North and away from us to the South of the
  galaxy disk.  In both cases we are probing the region of the NLR
  where the jet has ``broken-out'' of the galaxy disk.  The higher
  velocity gas is moving along the walls of the bicone, as accelerated
  by the jet \citep[e.g.,][]{storchibergmannetal2010,riffeletal2013}.

The coadded spectra from different regions reveal a similar
story. In the disk regions (regions 2 \& 3), where there is little
emission from the NLR gas, we find strong and narrow \hmol, and
weak, but rather broad, \brg\ and [\ion{Si}{6}].  In the NLR regions
(4 \& 5), \brg\ and \sivi\ are stronger and slightly broader than in
the disk regions.  The line profiles are most complex at positions
6, 7, and 8, regions that are also bright in \oiii\ emission and
where we see the strongest skewness in the lines
(Fig. \ref{fig:oiiisivi} \& Fig. \ref{fig:2dgas}). We clearly see
the red wing in the spectra from regions 6 \& 8, and the
corresponding blue wing in the spectra from region 7.

\includegraphics[width=0.45\textwidth]{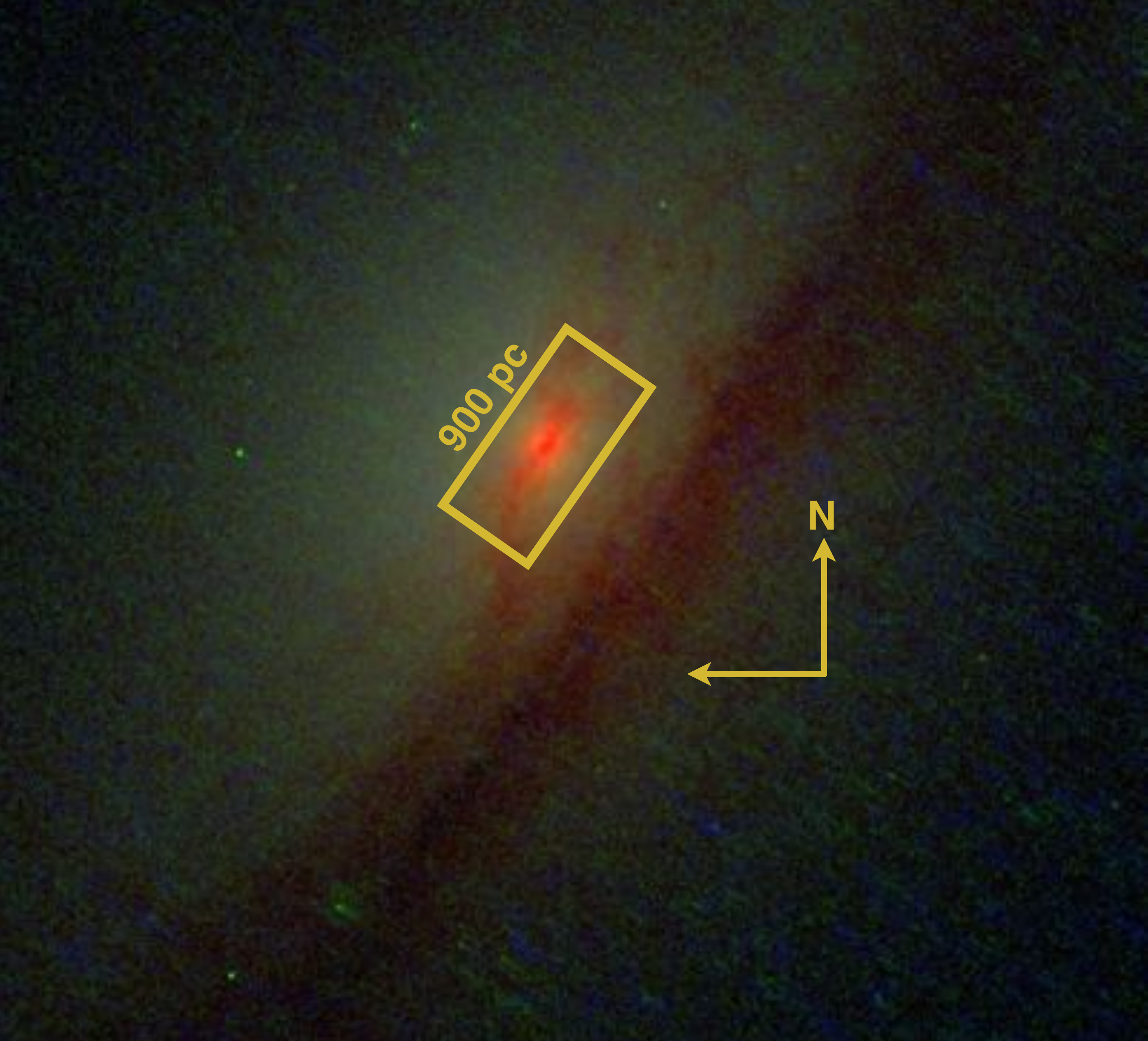}
\figcaption{Three-band \hst/WFC3 image of NGC 1194 including F336W, F438W, and F814W
  \citep{greeneetal2013}.  As indicated, North is up and East to the left 
  in this image. We show the OSIRIS field of view ($3.2 \times 1.2$\arcsec\ 
  or 900$\times$340 pc) 
  with the yellow box, oriented along the major axis of the galaxy as our 
  observation was.
\label{fig:1194image}
}
\vskip 5mm

Again in line with recent work \citep[e.g.,][]{storchibergmannetal2010}, we
find that the high-ionization lines and hydrogen recombination lines
trace ``feedback''.  That is, the emission in these lines is dominated
by extraplanar gas likely excited by the AGN, and perhaps entrained by
the nuclear jet.  This NLR emission is also aligned roughly
perpendicular to the megamaser disk on sub-pc scales.

Finally, we ask whether there is a broad component to the \brg, since
\citet{hfs1997broad} detect a weak broad base to the \halpha\ line in
NGC 4388, with a FWHM of 3900~\kms.  Within the central nuclear point
source emission, all the species have lines that are broad and
symmetric (Table 1).  However, we find no evidence for a true
broad-line component in \brg.  Instead, all the transitions have
comparable widths.  Neither \citet{veilleuxetal1997} nor
\citet{lutzetal2002} detect broad emission from \brg\ or Br$\alpha$ in
their long-slit NIR spectra either. Perhaps the broad component
detected in the optical is scattered broad emission rather than
directly transmitted light \citep[e.g.,][]{liuetal2009}.

\section{NGC 1194}

NGC 1194 is an inclined ($i \approx 90$\degr) S0 galaxy at $D=57.9$
Mpc.  The galaxy PA is 145\degr, and we aligned the IFU along the
major axis. Given our PSF of $\sim 0\farcs16$, we are resolving 40 pc
scales with our AO-assisted observations.  Thus we achieve very
comparable spatial resolution to the NGC 4388 observations presented
above. We have detected the
large-scale disk in NGC 1194 in \ion{H}{1}, and find a circular
velocity of $220 \pm 20$~\kms\ \citep{sunetal2013}.  From SDSS
imaging, we find a bulge-to-total ratio of $0.5 \pm 0.2$
\citep{greeneetal2010}.

The BH in NGC 1194 has a mass of $6.5 \pm 0.4 \times 10^7$~\msun, one
of the most massive in the megamaser disk sample.  NGC 1194 is in the
\emph{IRAS-}selected 12$~\micron$ sample \citep{rushetal1993}, and in
the SWIFT/BAT 22 month sample \citep{tuelleretal2010}.  It has a hard
X-ray luminosity of log $L_x = 43.2$ from 14-195 keV, but is not 
Compton thick \citep{georgantopoulosetal2011}.  Depending on
the bolometric correction, the Eddington ratio is $\approx
10^{-3}-10^{-2}$ \citep{vasudevanetal2009}, considerably lower than
NGC 4388.  The galaxy was also part of the \hst/\oiii\ snapshot survey
of \citet{schmittetal2003}.  They find extended \oiii\ emission with
an extent of 700 pc along the major axis of the galaxy, and 470 pc
along the minor axis.  The nucleus was detected in FIRST and at
8.4 GHz by \citet{theanetal2000}, with a flux density of 0.9 mJy, and
an upper limit on the radio core size of 52 pc.

\includegraphics[width=0.45\textwidth]{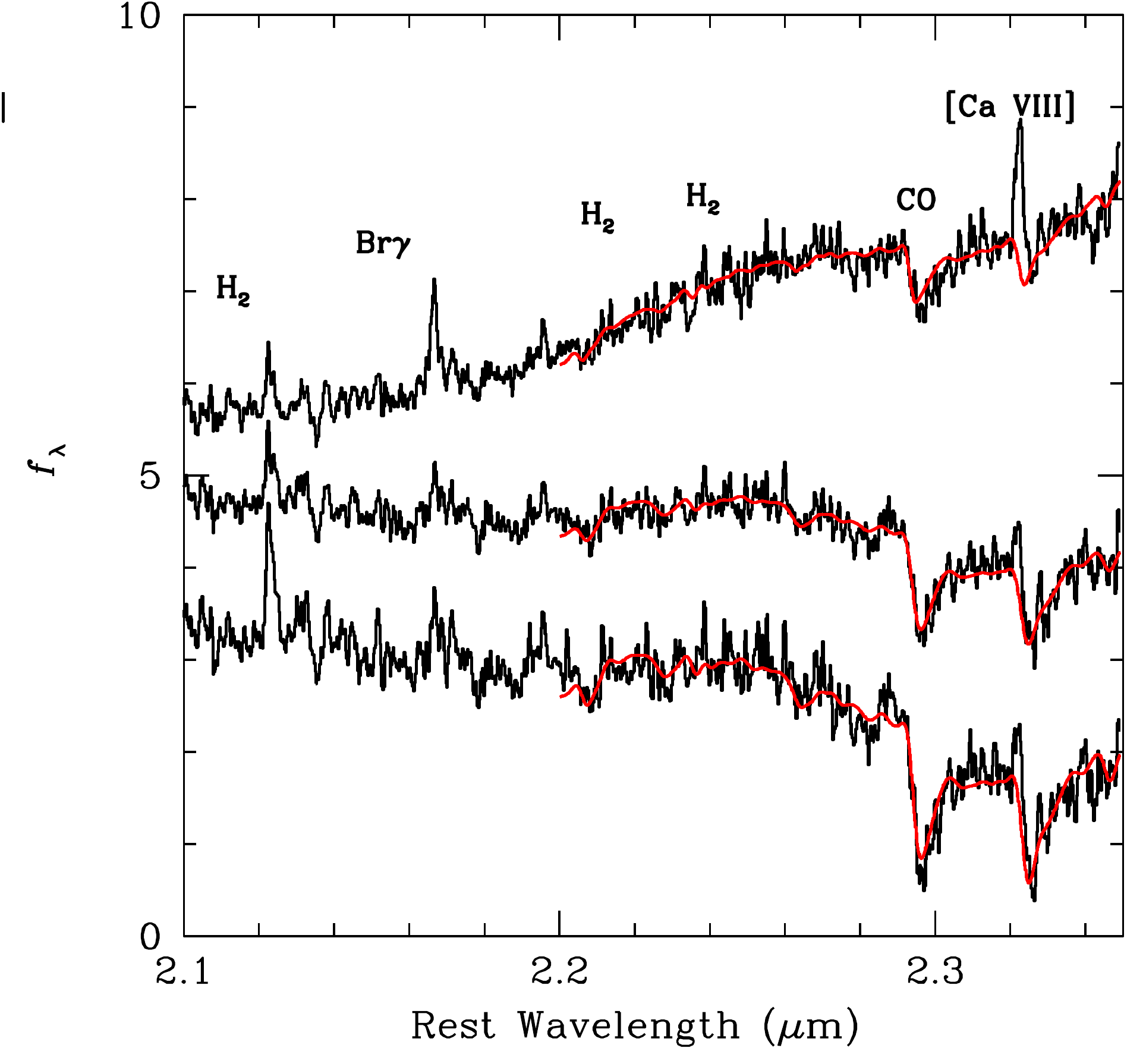}
\figcaption{Three spectra from the NGC 1194 data cube; due to telluric residuals 
  we do not show the region from $1.95-2.1$~~\micron. 
  From top to bottom, we show
  spectra within the inner 0\farcs18 (50 pc), between 0\farcs2 and 0\farcs5 
  (140 pc),
  and then 0\farcs5 to 1\arcsec\ (140-280 pc).  Going from the center outward, we
  can see the continuum change from AGN-dominated to galaxy-dominated, the
  \hmol\ equivalent width increase, and the [Ca {\tiny VIII}] line
  strength decrease.  These trends are qualitatively similar to NGC
  4388 above, although we have far less detailed spatial
  information. In red, we overplot our best-fit composite stellar template
  over the restricted region that we fit, and as above the strong emission lines 
  were masked in the fit.
\label{fig:lineregions}
}
\vskip 5mm

There are three existing $K-$band spectra of NGC 1194
\citep{sosabritoetal2001, imanishialonsoherrero2004,daviesetal2005} on
$\sim 1$\arcsec\ (900 pc) scales.  In all cases the CO bandhead is detected,
but the line emission is very weak. Davies et al.\ do not detect \brg,
but marginally detect the \hmol$~\lambda~2.12 ~\micron$ line.

\subsection{Stellar and Gas Kinematics}

\begin{figure*}
\includegraphics[width=0.9\textwidth]{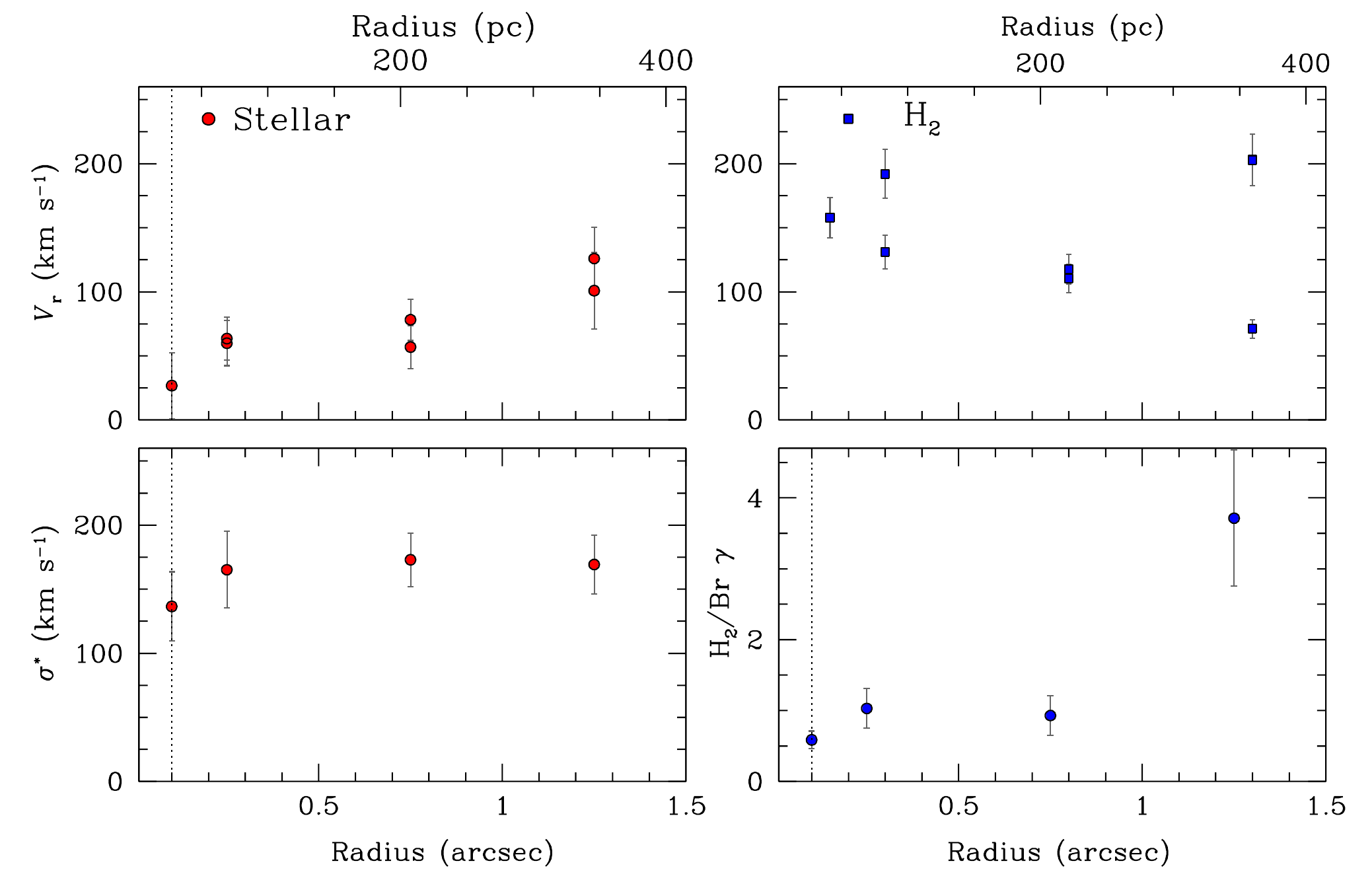}
\figcaption{The NGC 1194 data have considerably lower S/N than the 
NGC 4388 data.  We have binned
  the spectra into seven zones, including the nucleus within the central 
  $0\farcs16$, and then three annuli of width 0\farcs5 on either side 
  of the galaxy, extending along the major axis of the galaxy.  
  The velocity and dispersion are measured individually for each of these 
  bins. We examine the stellar and \hmol\
  rotation curves (top).  We also show the stellar velocity dispersion
  (bottom left) as a function of radius, measured from radial bins
  (e.g., folding about the major axis).  Finally, we show that the
  ratio of \hmol\ to \brg\ rises dramatically from the center where it
  is AGN-dominated to the disk at larger radius (bottom right).
\label{fig:n1194stelsigcurve}
}
\end{figure*}

Due to the heavy binning required to achieve adequate S/N, we cannot
derive much spatially resolved information for this galaxy.  In
  terms of the gas, the \brg\ and \sivi\ flux distributions are very
  centrally concentrated, with a FWHM that is consistent with that of
  the continuum. The \hmol$~\lambda 2.12 ~\micron$ emission is clearly
  extended, but the distribution is quite patchy. We therefore
construct average spectra in annuli, taking the central spectrum
within the FWHM of the PSF (0\farcs16). Then we construct six spectra
with outer radii of 0\farcs5, 1\arcsec, and 1\farcs5 (140, 280, and
420 pc) respectively, on either side of the galaxy minor axis.  Three
examples of these spectra are shown in Figure \ref{fig:lineregions}.

We can measure the radial velocities and stellar velocity dispersions
reliably in these coadded spectra, using pPXF as above.  We do find
that the stars are rotating, with an amplitude of $\sim 100$~\kms\
(Figure \ref{fig:n1194stelsigcurve}).  In contrast, we do not detect a
clear rotation curve in the gas.  Instead, the velocities are very
asymmetric. Asymmetric velocities in high-ionization lines are often
attributed to outflow
\citep[e.g.,][]{mullersanchezetal2011,mazzalayetal2013n1068}, and are
sometimes seen in \hmol\ as well
\citep[e.g.,][]{riffelstorchibergmann2011}, which may be the
explanation here.  Another possibility is that we are seeing spiral
arms or some other non-axisymmetric gas distribution that is unrelated
to the AGN. Our limited S/N prevents us from distinguishing between
these cases. Neither the \sivi\ nor the \brg\ is spatially extended
enough to measure reliable velocities beyond the central region.

\subsection{Gas Emission Lines}

All of the gas emission lines fluxes are weak, and the intrinsic
luminosities are also low.  It is perhaps not surprising that this S0
galaxy is gas poor compared to NGC 4388 above, based simply on their
respective morphology. On the other hand, the center of NGC 1194 is
quite dusty, so there must be some associated gas, and indeed we
detect atomic hydrogen in this galaxy on large scales
\citep{sunetal2013}.  The strongest gas emission comes from the
high-ionization [Si {\small VI}] line, but as mentioned above it is
only marginally spatially resolved.  Thus, we suspect that the
emission from this transition and \brg\ emerge from the inner NLR. In
contrast, the \hmol\ is clearly spatially resolved.  In analogy with
NGC 4388 above, these different gas morphologies may be suggesting
that the \hmol\ is excited by both the AGN and stellar processes. We
also measure the spatially resolved ratio of \hmol/\brg, and find that
the ratio is lowest in the galaxy center (having AGN-like values) and
then rises outward, perhaps indicating an increased contribution to
the \hmol\ from stellar processes.  These line ratios may suggest that
the asymmetric velocities we observe in the \hmol\ are due to
non-circular motions in the inner regions of the galaxy (e.g., spiral
arms) rather than outflow.

\section{Summary}

We have analyzed integral-field $K-$band observations of two megamaser
disk galaxies.  Our data probe $\sim 50$ pc scales in both galaxies,
and thus allow us to study the distributions of stars and gas at
the centers of these galaxies to investigate AGN fueling.
In NGC 1194, an S0, there is very little gas, and with our S/N ratios, 
we cannot say much about the two-dimensional velocity or 
dispersion fields.  However, in NGC 4388 we have excellent S/N, and 
uncover a variety of interesting features:
\begin{enumerate}

\item
The stellar velocity field demonstrates well-ordered rotation aligned with 
the kpc-scale disk.  Although there is net rotation in the stars, still they 
are kinematically hot, with $V/\sigma \approx 0.6$, likely dominated by 
the dispersion in the kpc-scale bulge. 

\item
  In the inner 1\arcsec\ (100 pc), we also see evidence for a nuclear
  disk, offet in PA by 15\degr\ from the kpc-scale disk. The evidence
  includes disky isophotes in the \hst/WFC3 F160W image and a distinct
  drop in stellar velocity dispersion on one side of the putative disk
  (the other side apparently extincted). Features like this, so called
  $\sigma-$drops, have been seen in the centers of a number of nearby
  spirals \citep[e.g.,][]{emsellemetal2001,peletieretal2007}, and there is some hint
  that they are more common among actively accreting galaxies
  \citep{hicksetal2013}.  We note that this 100 pc-scale nuclear disk
  is misaligned by $\sim$35\degr\ from the megamaser disk on sub-pc
  scales.

\item
  The \hmol\ gas also shows regular kinematics on $> 100$ pc scales,
  with a well-defined rotation field that is aligned with the
  kpc-scale disk. With a higher rotation amplitude, the gas is
  considerably colder than the stars ($V/\sigma \approx 1.5$). We also see
  a kink in the rotation curve on $\sim 100$ pc scales. Intriguingly,
  the \hmol\ rotation within 100 pc appears to align with the
  megamaser disk.  However, the physical origin of the kinematic twist
  is not yet clear. 

\item
Based on \hmol\ diagnostic line ratios, we conclude that 
the molecular hydrogen is mainly excited by thermal processes.
Based on the morphology and kinematics of the \hmol\ gas, we speculate that 
these thermal processes are mostly associated with stars rather than
radiation from the AGN \citep[e.g.,][]{storchibergmannetal2010,
riffelstorchibergmann2011}.

\item
  In contrast, the \brg\ and high-ionization lines (particularly [Si
  {\small VI}]) have a completely different spatial distribution and
  kinematics.  They trace the inner narrow-line region, also seen on
  500 pc scales in \oiii, as well as the 100 pc-scale jet.

\end{enumerate}

In the future, we hope to analyze a larger sample of megamaser disk
galaxies in this manner. Combined with direct tracers of the cold
molecular gas (e.g., with ALMA), as well as high-resolution imaging
from \hst, we hope to build a multi-phase map of AGN fueling from 100
to 0.1 pc \citep{greeneetal2013}.

\acknowledgements 
The referee provided a very thorough review that improved the 
quality of this manuscript. 
We thank K. Gebhardt and R. van den Bosch for assistance with the
early stages of this project.  J.L.W. has been supported by an NSF
Astronomy and Astrophysics Postdoctoral Fellowship under Award
No. 1102845. Some of the data presented herein were obtained at the
W.M. Keck Observatory, which is operated as a scientific partnership
among the California Institute of Technology, the University of
California and the National Aeronautics and Space Administration. The
Observatory was made possible by the generous financial support of the
W.M. Keck Foundation.  The authors wish to recognize and acknowledge
the very significant cultural role and reverence that the summit of
Mauna Kea has always had within the indigenous Hawaiian community.  We
are most fortunate to have the opportunity to conduct observations
from this mountain.  Keck telescope time was granted by NOAO, through
the Telescope System Instrumentation Program (TSIP). TSIP is funded by
NSF.


\end{document}